\documentclass[12pt]{article}
\usepackage[utf8]{inputenc}
\usepackage{graphicx}
\usepackage{subfig}
\usepackage{setspace}
\usepackage{mathtools}
\usepackage{amsmath}
\usepackage{nccmath}
\usepackage{pdfpages}
\usepackage{cite}
\usepackage{caption}
\usepackage{placeins}
\usepackage[margin=1in]{geometry}
\usepackage{hyperref}
\usepackage{placeins}

\title{Does the Use of Unusual Combinations of Datasets
Contribute to Greater Scientific Impact?}
\author{
    Yulin Yu\textsuperscript{*1} ,  Daniel M. Romero\textsuperscript{1,2} \\
    \textsuperscript{1} School of Information, University of Michigan \\
    \textsuperscript{2} Center for the Study of Complex Systems, University of Michigan \\
     \textsuperscript{*} Address correspondence to: yulinyu@umich.edu\\
}

\begin{document}
\maketitle

\section*{Abstract}
Scientific datasets play a crucial role in contemporary data-driven research, as they allow for the progress of science by facilitating the discovery of new patterns and phenomena. 
This mounting demand for empirical research raises important questions on how strategic data utilization in research projects can stimulate scientific advancement. In this study, we examine the hypothesis inspired by the recombination theory, which suggests that innovative combinations of existing knowledge, including the use of unusual combinations of datasets, can lead to high-impact discoveries. Focusing on social science, we investigate the scientific outcomes of such atypical data combinations in more than 30,000 publications that leverage over 5,000 datasets curated within one of the largest social science databases, ICPSR. This study offers four important insights. First, combining datasets, particularly those infrequently paired, significantly contributes to both scientific and broader impacts (e.g., dissemination to the general public). Second, infrequently paired datasets maintain a strong association with citation even after controlling for the atypicality of dataset topics. In contrast, the atypicality of dataset topics has a much smaller positive impact on citation counts. Third, smaller and less experienced research teams tend to use atypical combinations of datasets in research more frequently than their larger and more experienced counterparts. Lastly, despite the benefits of data combination, papers that amalgamate data remain infrequent. This finding suggests that the unconventional combination of datasets is an under-utilized but powerful strategy correlated with the scientific impact and broader dissemination of scientific discoveries.

\section*{Introduction}

The recognition of the immense power of datasets in scientific and economic advancements has prompted academia, industry, and society to collectively invest substantial effort in generating and making datasets publicly available ~\cite{Mayer-Schonberger2013-sx,Gewin2016-cp,Hanson2011-ge,piwowar2013data}. The open science movement, for instance, has emphasized the crucial practice of data sharing to enhance research reproducibility, facilitate collaboration, and enable subsequent studies~\cite{Hossain2016-fy,Murray-Rust2008-jl,Scheffler2022-ev,Milham2018-ln}. Initially, the call for sharing and managing datasets faced numerous barriers, including limited funding~\cite{tenopir2011data}, inadequate institutional support, time constraints~\cite{Tenopir2020-um}, lack of suitable platforms ~\cite{Markiewicz2021-ft}, and a lack of sharing social norms in academia~\cite{Andreoli-Versbach2014-qw, Hipsley2019-hs}. Fortunately, over the past decade, there has been a notable increase in funding, institutional support, and the development of platforms dedicated to supporting data sharing and curation.\cite{Kaiser2023-dz,devriendt2020data,mendez2020progress}. As a result, a wealth of publicly available datasets is now accessible for reuse~\cite{Wicherts2011-sd,Kaiser2023-dz,Kelling2009-dv,Fecher2015-qn,Hill2020-wo}.

Given the wide accessibility of publicly available data and the significance of datasets within the scientific community, it is crucial to understand \emph{how} scientists utilize these datasets, especially when their use fosters high-impact and innovative scientific development. Numerous studies have endeavored to discern the motivations and challenges surrounding the reuse of datasets~\cite{pasquetto2017reuse,pasquetto2019uses,tenopir2011data}. These studies aim to promote data reuse and enhance the curation process (such as improving the data search experience)~\cite{lafia2023direct,viswanathan-etal-2023-datafinder}, thereby encouraging researchers to effectively utilize existing datasets or identify suitable data for their studies. However, the link between dataset utilization and scientific advancement remains uncertain. In this study, we aim to fill this gap, particularly by analyzing how strategic data utilization in research projects can drive scientific advancement and foster high-impact, innovative scientific development. 

A line of study in recombination theory offers a broader perspective on the potential relationship between diversity (unusual combinations) and scientific advancement. This body of literature suggests that unconventional combinations of existing knowledge that retain a certain level of conventionality (e.g., combining two high-impact findings from different domains) can lead to novel discoveries and scientific breakthroughs~\cite{Leahey_undated-mz,Uzzi2013-va,Shi2023-xv,Foster2015-bm,Lin2022-pa,Leahey_undated-mz}. While these studies do not offer
empirical evidence linking data usage practices to scientific advancement, they do provide valuable insights that lead us to ask: do unconventional combinations of datasets contribute to scientific breakthroughs? Examining the scientific impact of novel dataset combinations can provide valuable insights for publication agencies, data curators, researchers, and funders. These findings can inform the development of policies and practices that facilitate and encourage data linking, ultimately enhancing the overall research landscape. 

Using Social Science as a testbed for our analysis, we begin by examining whether papers that use multiple datasets receive more scientific attention (citations and mentions on various online platforms) than their counterparts that only use one dataset. We then assess whether papers employing an atypical combination of datasets receive more scientific attention. Moreover, we leverage the topic tags linked to each individual dataset to measure a paper's uniqueness in combining datasets of different topics. We aim to explore whether novelty in dataset combinations and topic combinations exerts distinct influences on scientific attention.

We have compiled a comprehensive dataset comprising more than 30,000 papers that utilize over 5,000 distinct datasets. The dataset used in our study was obtained from the Inter-university Consortium for Political and Social Research (ICPSR)\footnote{url{https://www.icpsr.umich.edu/web/pages/}}, a renowned data curation service extensively utilized by social scientists. This dataset is meticulously labeled by ICPSR data curators, and the linkage between datasets and publications is established only when a paper extensively employs a particular dataset in its results~\cite{hemphill2024dataset,lafia2022subdivisions}. The precision of data usage within publications is crucial for our analysis due to two reasons: first, data citations are commonly absent from certain publications~\cite{fenner2019data,cousijn2019bringing}, and second, datasets are often cited for various reasons, many of which do not indicate substantial reuse (e.g., tangentially citing in the introduction or discussion)~\cite{moss2018opaque,blake2010beyond}. After identifying this dataset from ICPSR, we connect the publication record data through OpenAlex\footnote{url{https://openalex.org/}}, and Altmetric\footnote{url{https://www.altmetric.com/}}, which provide information on citations and mentions of the research papers over the past decade on online platforms, such as news and social media. A complete description of our data is provided in the 'Materials and Methods' section and in the Supplemental Information (SI) Appendix, section 1. We investigate the impact of unique data combinations across multiple dimensions, including scientific impact (e.g., citations) and broader dissemination (e.g., news mentions, policy implications, and general knowledge impact). Consequently, our study offers a systematic investigation into the effect of the uniqueness of data integration on scientific impact and broader dissemination.

\subsection*{The effect of dataset combinations on scientific impact}

A prerequisite for data combination is using multiple datasets. Thus, our analysis begins by examining the impact of using multiple datasets on the paper's citations. Our primary citation impact metric is the number of citations a paper obtained in the fixed number of years after publication. 

Since our outcome variable tracks the counts of citations and exhibits a long tail distribution (see Figure S2 in Supplemental Information (SI)), we employ a Negative Binomial regression to effectively model the relationship between citation count and the use of multiple datasets. In our dataset, 30,479 papers use only one dataset, while 8,836 papers use multiple datasets. Among these, 3,497 papers use exactly two datasets (Refer to Figure S1 in the Supplemental Information (SI) for the distribution of the number of datasets.
). Our model controls for several variables. First, since the frequency of a dataset’s use in other papers could signal its relevance to a popular topic or its scientific value and lead to more citations of the focal paper, we control for dataset usage frequency. While citations to a dataset paper   (i.e., a paper contributing the new data set) would also be a natural control variable to consider \cite{mcgillivray2022deep, funk2017dynamic}, we do not include this control because the majority of ICPSR datasets are standalone and not connected to a dataset paper. Additionally, we control for the impact of team size, team experience, disciplines, publication time, and journal impact factor, as these factors have been found to influence citation performance due to their influence on team composition and journal characteristics~\cite{wuchty2007increasing, mccain1990mapping, radicchi2008universality, redner1998popular}. We provide a detailed explanation of these measurements in the Supplemental Information (SI) Appendix Section 3.

In our initial analysis, we use a binary variable that encodes whether a publication utilizes multiple datasets (data combination) or a single dataset and the number of citations 3, 5, and 10 years after publication as the outcome variable. As shown in Figure 1, the Negative Binomial regression results show a statistically significant increase in citations for papers that use multiple datasets, compared to those that did not (P-value $<$ 0.001). Papers that used more than one dataset garnered 17.1\%, 15.2\%, and 14.0\% more citations over 3, 5, and 10 years relative to papers that used a single dataset (See Supplemental Information (SI) Appendix Section 4, Regression Tables S1 - S3). As shown in the inset of Figure 1, the effect has remained significant and consistent over time, except for papers published before 1900, when our data becomes very sparse. There is a notably larger effect size in recent years, particularly after 1990 (see SI Appendix section 4 Tables S4 - S7 for full regression tables).

\begin{figure}
\centering
\includegraphics[width=.8\linewidth]{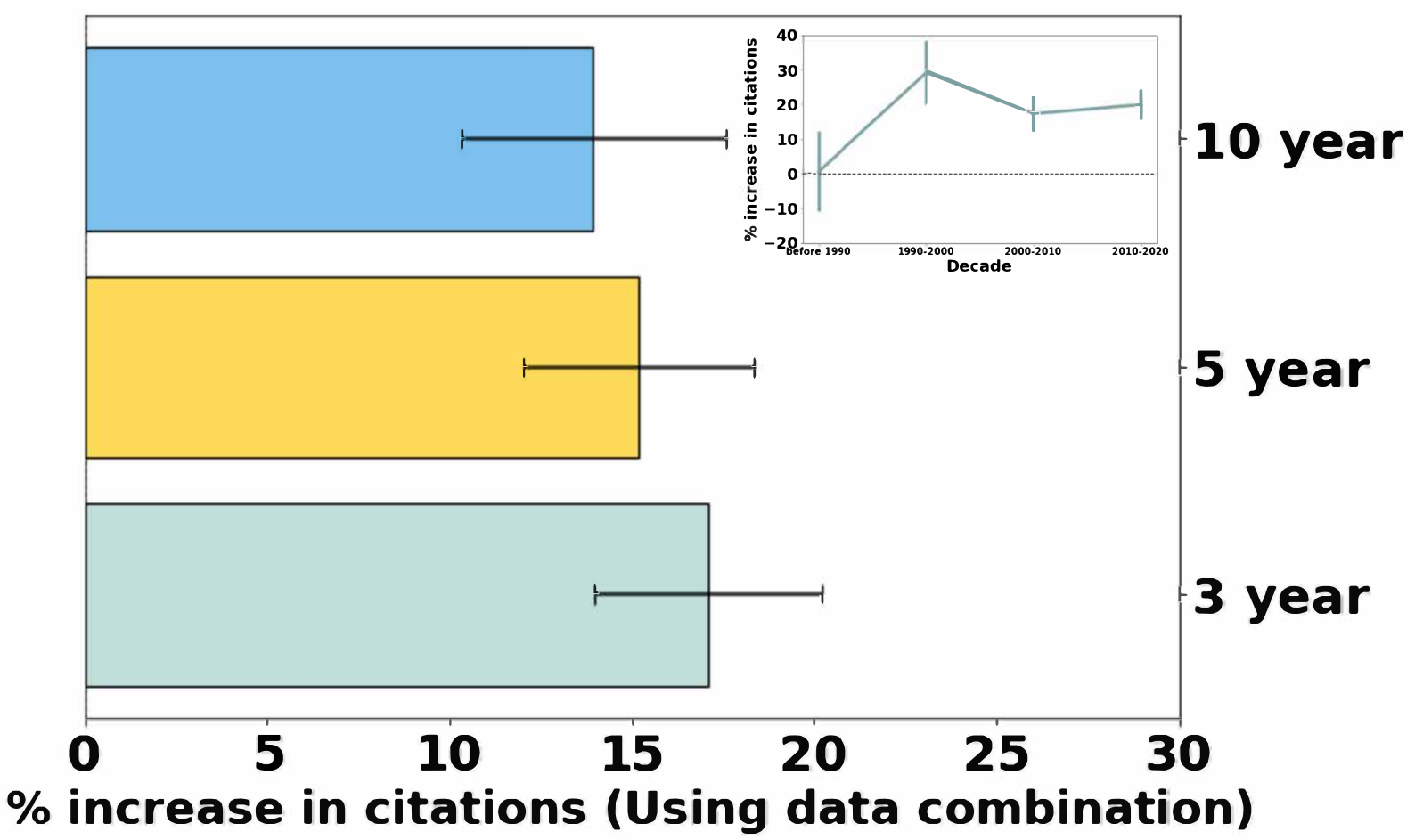}
\caption{The plot illustrates the effect size of the data combination variable on citations over 3, 5, and 10 years based on a Negative Binomial regression. The error bars indicate the 95\% confidence intervals. This regression controls for dataset usage frequency, author attributes (i.e., the number of authors and their experience), journal impact factor, publication year, and subject areas. The results show a positive effect of data combination on citations over 3, 5, and 10 years. The inset of the figure shows the effect size and 95\% confidence intervals for analyses conducted on publications published in four distinct periods: before 1990, 1990-2000, 2000-2010, and 2010-2020. The results reveal that the effect of data combination on citations is driven by papers published in more recent years, particularly those after 2000.}
\label{fig:a1}
\end{figure}

We conducted additional robustness tests and found consistent results, including 1) performing a regression analysis treating our binary data combination variable as a continuous one representing the number of datasets used in a paper. The results show that using an additional dataset in a paper is associated with 1.4\%, 1.0\%, and 0.8\%  more citations over 3, 5, and 10 years, respectively (See Supplemental Information (SI) Appendix Section 4, Regression Tables S8 - S10), 2) applying a model-free robustness test comparing the number of citations of papers using multiple datasets with a matched set of papers using a single dataset. (see Supplemental Information (SI) Appendix, Section 6 for details)

\subsection*{Atypical combinations of datasets are associated with high impact}
Subsequently, we evaluate the impact of data combination beyond the number of datasets; we now consider the impact of using rarely combined datasets. We assess the atypicality of a paper's data usage by employing the Rao-Stirling index~\cite{cassi2017analysing,Stirling2007-ya}, a general measure of atypicality. Prior studies have utilized the Rao-Stirling index to quantify atypicality in combining references or multidisciplinarity~\cite{Yang2022-zp,porter2009science}. 
Applied to our data, the Rao-Stirling index varies from 0.25 to 0.92 (Refer to Figure S3 in the
Supplemental Information (SI) for the distribution of the atypicality of datasets.), with higher values denoting greater atypicality. The 'Materials and Methods' section provides additional operational details of this measure. We also provide examples representing the top 25\% and bottom 25\% of data combinations, as determined by the atypicality of data combination score, in the Supplemental
Information (SI) Appendix, section 2. Figure 2(C) also illustrates the measurement.

We employed fixed-effects Negative Binomial regressions to investigate the effect of the atypicality of dataset combination on the citation impact of a paper. In the primary analysis, we examine exclusively 8,836 papers that employ a minimum of two datasets and utilize a three-year citation window to determine the citation impact of publications. We also use control variables, including team size, team experience, journal impact factor, disciplines, publication time, and average data use frequency, as in the previous analysis. Additionally, we control for the number of datasets utilized in the paper, which have been shown to be associated with citation impact in our first analysis. Furthermore, acknowledging prior research that show a link between atypical combinations of prior knowledge and citations~\cite{uzzi2013atypical,Foster2015-bm}, we control for atypicality in the combination of journals referenced by the paper, which we refer to as \emph{paper novelty} in our study (See Materials and Methods for a full description of this measure). 

Figure 2(A) displays the regression coefficients of the independent and several control variables (see the full regression table in the Supplemental Information (SI) Appendix, section 4 Table S11.). The result suggests that papers that utilize more uncommonly combined datasets significantly garner more citations (P-value $<$0.001). For each standard deviation increase in data atypicality (corresponding to a 0.13 increase in the Rao-Stirling index), papers receive 18.4\% more citations obtained within 3 years after publication. These results are robust to different choices of citation time-window, including five (19.1\%) and ten years (19.9\%) after publication, as shown in the SI Appendix, section 4 Tables S11-S13.

The effect remains significant and consistent over time, with a noticeably larger effect size in recent years. As shown in the SI Appendix, section 4 Tables S14 - S17, the effects for each time period are as follows: before 1990 (9.6\%), 1990-2000(15.8\%), 2000-2010(21.3\%) and 2010-2020(22.2\%)

Similar to the prior analysis, we also conducted three additional robustness tests and found consistent results: 1) using a binary outcome variable indicating whether a paper is in the top 5\% most cited in our dataset ~\cite{uzzi2013atypical,Yang2022-zp} (refer to Supplemental Information (SI) Appendix, Sections 4, Regression Table S22), 2) applying a model-free robustness test comparing the number of citations of high atypicality papers with a matched set of low atypicality papers (see Supplemental Information (SI) Appendix Section 6 for details).
\begin{figure*}
\centering
\includegraphics[width=.95\linewidth]{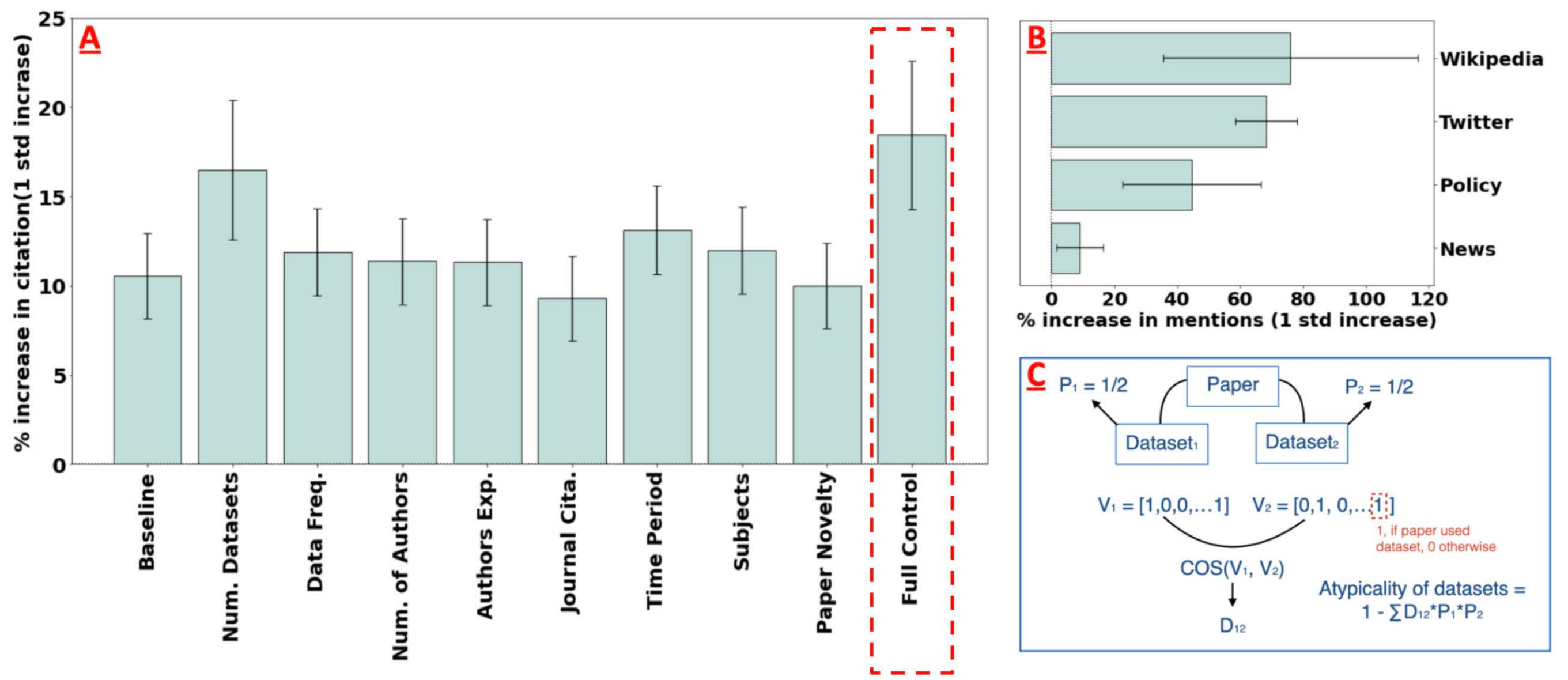}
\caption{Atypical combinations of datasets lead to higher citation rates and broader dissemination.  
(A) The plot illustrates the effect size of the atypicality in dataset combinations variable on citations over 3 years based on a Negative Binomial regression controlling for the various factors indicated in the panel headings. The leftmost panels display effect size of atypicality in baseline regressions (without any control variables), while the rightmost panels display effect size after collectively controlling for dataset attributes (use frequency and number of datasets), author attributes (number of authors and experience), journal impact factor, publication year, subjects, and paper novelty. The error bars indicate the 95\% confidence intervals.  (B) The effect size and 95\% confidence intervals provide insights into the impact of atypicality in dataset combination on Twitter, Wikipedia, policy, and news mentions (outcome variables) based on a Negative Binomial regression. This regression incorporates all of the control variables listed above. (C) Illustration of quantifying atypicality of datasets using the Rao-Stirling index. 
In this illustration, we assume that a paper uses two datasets, namely $dataset_1$ and $dataset_2$. We first vectorize each dataset into a one-hot vector. Each coordinate in the vector corresponds to a paper in our dataset, and the coordinate takes a value of 1 if the respective dataset is used in that paper and a value of 0 if otherwise. Subsequently, we calculate the distance, denoted as $D_{12}$, by computing cosine similarity between $dataset1$ and $dataset2$. Using the number of datasets in a given paper, we calculate the parameters $P_1$ and $P_2$, representing the ratio of a dataset used within a research paper. In this particular scenario, where two datasets are used in the paper, both $P_1$ and $P_2$ are equal to 1/2. We then use the equation shown in the figure to quantify the atypicality of the datasets used by the paper.} 
\label{fig:a1}
\end{figure*}
\section*{The effect of dataset combinations on scientific impact}
Furthermore, our analysis reveals that the impact of atypical dataset combinations on research publications extends beyond citation counts. We examine the effects of dataset atypicality on the broader online dissemination of research findings, including their presence in general knowledge platforms such as Wikipedia, their presence on policy documents, and their attention on social media (Twitter) and news platforms using fixed-effects Negative Binomial regressions with the control variables mentioned above. For this analysis, we consider papers published after 2010 because Altmetric, our data source for scientific online dissemination, began collecting data around this time.  As illustrated in Figure 2(B), a one standard deviation increase in atypical dataset combinations is associated with a 44.6\% increase in policy mentions, a 76.0\% increase in Wikipedia mentions, a 68.2\% increase in Twitter mentions, and a 9.0\% increase in news mentions (see the full regression tables in Supplemental Information (SI) Appendix, section 4 Tables S18 - S21.). A description of policy, Wikipedia, Twitter, and news mentions is provided in the Materials and Methods and the Supplemental
Information (SI) Appendix, section 1. 

\section*{The effect of atypical dataset topic combinations on scientific impact}

Does innovation arise from combining previously uncombined datasets or from combining datasets with unique \emph{topics}? Which approach has a stronger effect? The datasets in our analysis are tagged with a set of expert-defined topics. For instance, the dataset titled ``Cost of Living in the United States" includes topics such as ``consumers", ``the cost of living", ``economic indicators", ``expenses", ``families", ``households", ``income", ``urban populations", and ``the working class". This enables us to assess the atypicality, not only of individual datasets used in a paper but also of the atypicality of the topics covered by such datasets. Our subsequent analysis explores 1) if the effect of dataset atypicality holds after controlling for the atypicality of \emph{topics within datasets} and 2) the effect of both types of data combination on citation counts. 

\begin{figure*}
\centering
\includegraphics[width=.95\linewidth]{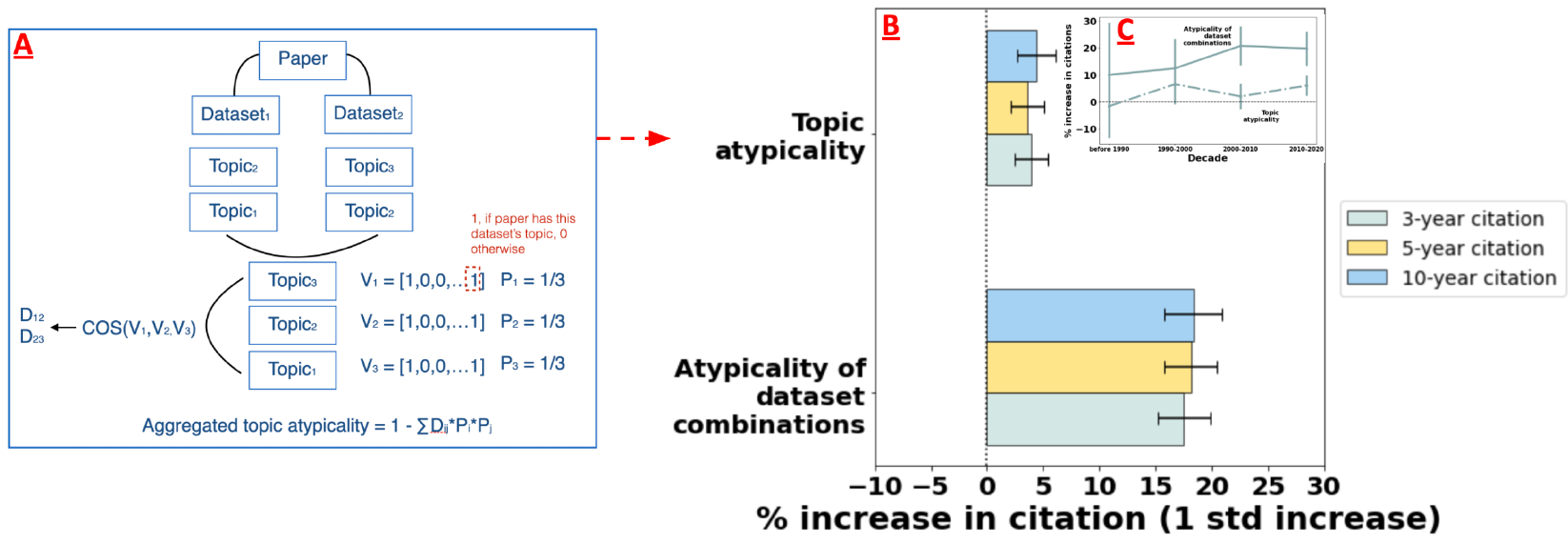}
\caption{A paper's atypical combination of datasets is more impactful than atypical combination of datasets' topics. (A) Illustration of quantifying topic atypicality: In this illustration, we consider a hypothetical paper that utilizes two datasets, namely $dataset_1$ and $dataset_2$. The first dataset, $dataset_1$, is associated with two topic tags: $Topic_1$ and $Topic_2$, while the second dataset, $dataset_2$, is associated with two topic tags: $Topic_2$ and $Topic_3$. We combine all the topics from both datasets, resulting in a topic set containing $Topic_1$, $Topic_2$, and $Topic_3$. Subsequently, we represent each of these topics as a one-hot vector. In this representation, each coordinate in the vector corresponds to a paper, and the coordinate takes a value of 1 if the respective topic is present in that paper; otherwise, it takes a value of 0. Using cosine similarity, we calculate the distance between these topic vectors, and we apply similar quantification methods to all pairs of topics. This process allows us to determine the topic atypicality within the paper. (B) The effect size of topic atypicality and atypicality of data combinations variables on citations over 3, 5, and 10 years based on a Negative Binomial regression. The error bars indicate the 95\% confidence intervals. Our model includes two main independent variables: topic atypicality and atypicality of data combinations. The dependent variable is the three, five, and ten-year citation counts, and the model incorporates all the control variables described in the preceding section (full control setting). (C) The effect size and 95\% confidence intervals are presented separately for analyses conducted on publications published in four distinct periods: before 1990, 1990-2000, 2000-2010, and 2010-2020. Our results reveal that effects are most influenced by the more recent years, particularly those after 2000. }
\label{fig:a1}
\end{figure*}

We measure topic atypicality in the datasets used by academic papers using the Rao-Stirling index described above, but considering the topics of the datasets, not the datasets themselves, as the units of analysis to be combined. We operationalize the metric for ``topic atypicality" by first taking the union of all topics associated with each dataset used by a paper and then measuring the atypicality of these topics. This determines whether the paper uses datasets that collectively combine atypical topics. We include examples representing the top 25\% and bottom 25\% topic atypicality, as determined by topic atypicality score, in the Supplemental
Information (SI) Appendix, section 2. Figure 3(A) illustrates the topic atypicality measure. In this analysis, we examine the 8,812 papers that employ at least two datasets and include datasets linked to at least two topic tags.

Figure 3(B) displays the regression results illustrating the correlation between citation impact and our two atypicality metrics: atypicality of dataset combinations and topic atypicality. We present these results after controlling for the variables mentioned in the preceding analysis. Our results show that even controlling topic atypicality, atypicality of dataset combinations still exhibits a strong and positive association with 3 (17.5\% increase), 5 (18.1\% increase), and 10 years (18.4\% increase) citations. This result remains robust across various settings and controls, including using  the hit paper variable used in our previous analysis as the dependent variable. Comparably, topic atypicality has a small and positive association with citations 3 (4.0\% increase), 5 (3.6\% increase), or 10 years (4.4\% increase) after publication. However, the effect of topic atypicality is not significant when using a binary outcome variable indicating whether a paper
is in the top 5\% most cited in our dataset (see the full regression tables in Supplemental Information (SI) Appendix, section 4 Tables S23-25/S30.). This suggests that 1) the use of uncommonly combined datasets dominates the effect on citations, regardless of whether the topic combination is novel and 2) the effect of using datasets with uncommonly combined topics might only have a slightly positive impact on citations.

Figure 3(C) displays separate analyses conducted for each decade: publications before 1990 (grouped due to limited observations), 1990-2000, 2000-2010, and 2010-2020. Our results reveal that the effect of atypical dataset combinations is driven by recent publications, particularly after 2000. While the effect is consistently positive over time, its size is significantly larger for publications in the last two decades than for papers from before the 21st century. The influence of topic atypicality is either very small or statistically insignificant (see the full regression table in Supplemental Information (SI) Appendix, section 4 Table S26 -S29.).

\section*{What type of research teams combine atypical datasets?}

Given that data combinations, particularly atypical combinations, contribute to scientific impact, our final analysis aims to understand which research teams are more likely to employ such datasets. 
Prior studies have emphasized the significance of teams in fostering scientific innovation~\cite{uzzi2013atypical,wu2019large} and have focused especially on team size. It is also known that the age or experience of authors is associated with creativity and innovation~\cite{jones2009burden}. We utilize logistic regression to model two relationships: 1) the likelihood of using multiple datasets (data combination) and team size (Model A in Fig.4) and 2) the likelihood of using multiple datasets and team experience as measured by the average number of citation counts of authors(Model C). Then, we use Ordinary Least Squares (OLS) regression to model relationships between:  3) team size and the atypicality of data combinations (Model B), and 4) the atypicality of data combinations and team experience (Model D). In models 1) and 2), we control for the average data use frequency and the impact factor of the journal. In models 3) and 4), we further include control variables for the number of datasets. We find that larger teams tend to use multiple datasets. Furthermore, smaller or less experienced teams tend to use atypical combinations of datasets. However, when examining all the research teams in our dataset, we observe that less than 30\% of the papers (29\%) in our analysis incorporate multiple datasets (See the Supplemental
Information (SI) Appendix, section 4 Table S31 - S34, for full regression tables).

\begin{figure*}
\centering
\includegraphics[width=.75\linewidth]{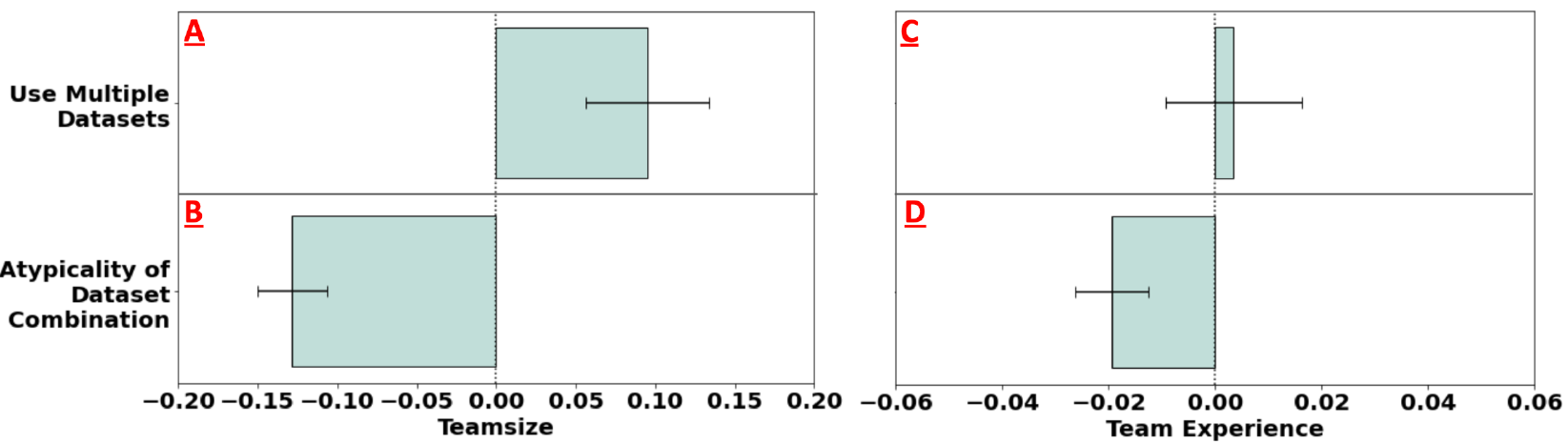}
\caption{While large team are more likely to use multiple datasets (data combination), smaller and less experienced teams are more inclined to use atypical dataset combinations. (A)(B) Impact of Team Size on Dataset Utilization: The regression coefficient and 95\% confidence intervals reveal the effect of the number of authors (team size) on the likelihood of research teams utilizing multiple datasets and incorporating atypical combinations of datasets in their papers. (C)(D) Impact of Team Experience on Dataset Utilization: The regression coefficient and 95\% confidence intervals demonstrate the effect of team experience, measured by the average number of citation counts of authors, on the likelihood of research teams utilizing multiple datasets and incorporating atypical combinations of datasets in their papers.
}
\label{fig:a1_update}
\end{figure*}

\section*{Discussion}

By conducting a meticulous analysis of a curated dataset comprising over 30,000 papers and over 5,000 datasets, we have found that the combination of datasets, particularly those that are not typically combined, is associated with an overall higher citation rate (an 18.4\% increase per one standard deviation difference). This finding remains robust after controlling for various factors, such as disciplines, team compositions, time periods, and paper novelty. We also find that atypical combinations of datasets have a significantly greater impact on citations than those employing datasets with unconventional topics. Regardless of how novel the dataset topics are, using atypical combinations of datasets exhibits a strong association with increased citations.

Our study parallels previous research asserting that atypical combinations of knowledge lead to high-impact scientific findings~\cite{uzzi2013atypical}. However, we build upon this work by uncovering specific implications for data use. While previous studies have examined novelty in a uni-dimensional manner, without specifying how various aspects of a scientific paper's novelty relate to its scientific impact~\cite{Boudreau2012-rq, Verhoeven2016-cf,Wang2017-fs,Lee2015-lb,Wu2019-gf,Uzzi2013-va,Bornmann2019-ok,Shibayama2021-ds}, or have solely investigated novelty in certain aspects that are orthogonal to data use (such as methods, theories, and finding) ~\cite{leahey2023types}, this research advances our understanding on the association between data use and impact.

Importantly, we found that novelty in datasets, the cornerstone of data-driven studies, has a larger effect size on citations than paper novelty as measured by references. This suggests that novelty in dataset use may be as impactful as the novelty of knowledge \cite{uzzi2013atypical}. Our results suggest that integrating not only novel combinations of knowledge but also novel combinations of empirical evidence (data) has strong benefits to the recognition of scientific work. Our results show that the benefit of atypical data combinations is not fully explained by the exploration of novel topic combinations, suggesting that employing novel combinations of datasets to study conventional topic combinations enhances scientific impact. This effect could be driven by the discovery of datasets that, when combined, fill an empirical gap and enable researchers to answer conventional research questions. However, further quantitative and qualitative research is necessary to fully understand the mechanisms behind this effect. 

Beyond academic contributions, our findings also present strategies for scientists, policymakers, and data curators for using and managing scientific data for research. We encourage researchers to explore new empirical resources by combining infrequently paired datasets. Given that data combination has a significant effect on producing high-impact scientific findings, policymakers may encourage or even require the publication of data, particularly when it includes the possibility of linking to other data sources~\cite{Hill2020-wo,Heberling2021-ge}. Similarly, data curators might consider making datasets more ``linkable" to other public datasets. For example, in research that employs individual publications as observations, we recommend that researchers include DOIs in their published data to facilitate linking by other researchers. Concurrently, data curators could create a data recommender system that considers the novelty of data pairings as part of the recommendations for data use.~\cite{viswanathan-etal-2023-datafinder}.

Our findings are not without limitations. First, while ICPSR is arguably the most suitable data source for this study, being ``the world’s largest archive of digital social science data" according to its website~\footnote{\url{https://www.icpsr.umich.edu/}}, it is not comprehensive nor necessarily a representative sample of all datasets. The criteria for data inclusion at ICPSR are not clearly defined on a broader scale, even though its datasets are widely recognized and utilized in social science research. Additionally, our study requires us to control for various variables, leading us to exclude papers where such variables are inaccessible. This selection process could introduce selection biases along unobservable factors like country, institution, or the motivations behind data inclusion in ICPSR. Second, Our study presented citation counts over 3, 5, and 10-year windows. A large increase in the percentage of citations may not necessarily correspond to a large increase in the \emph{number} of citations, particularly in fields with smaller overall citation counts and short time periods, such as 3 years~\cite{wang2013citation}. Third, while p-values might not be particularly meaningful given our large sample size in addition to statistical significance, our main analysis shows an important effect size, with over a 15\% increase in citation difference observed in our main findings, including the association between citation counts and whether a paper uses multiple datasets (data combination) and the atypicality of data combination.

Several potential avenues for future research can be pursued based on the current results. First, considering the significant value of data combination in scientific outcomes, it is noteworthy that researchers seldom engage in this practice. Future studies could delve into the multiple stakeholders involved in data curation and research, aiming to understand the reasons behind the infrequent combination of data and the challenges encountered when attempting to combine datasets atypically. A comprehensive qualitative investigation may be necessary to shed light on these issues.

Second, we recommend a causal analysis to discern the mechanisms behind the increased scientific impact and broader dissemination resulting from the combination of usually paired data. For instance, does the scientific community place greater value on evaluating hypotheses across multiple datasets and different settings? Or does connecting data lead to evaluating new hypotheses that could not be tested before? 

Third, it is worth noting that our analysis focused solely on social science research. Future research should aim to replicate our findings in other disciplines. Moreover, exploring the potential heterogeneous effects of data combination across different fields, given disciplinary variations in data curation and usage practices~\cite{Tenopir2020-um}, would be valuable. However, the successful execution of such studies is contingent upon resolving the challenges associated with data citation infrastructure, such as the labor-intensive nature of manual data citation. Currently, even data journals are indexed differently in major scholarly databases, and their indexing is often incomplete~\cite{jiao2023exclusively}. There are also disciplinary differences in data citation practices~\cite{gregory2023tracing}.  We call for not only data collection in other disciplines regarding precise data citation but also complete and unified indexing and the adoption of standardized multidisciplinary data citation practices.

The strategic utilization of datasets in research holds promise for scientific advancement. Although combining datasets, particularly through atypical combinations, is not yet a common practice, our research suggests that promoting this approach among researchers, policies, and data curators could lead to scientific products that advance knowledge and raise awareness of scientific contributions.

\section*{Methods}
Our analysis is based on a dataset comprising over 30,479 papers published in the past six decades, which extensively utilize 5,241 datasets from the Inter-university Consortium for Political and Social Research (ICPSR)~\cite{hemphill2024dataset}. The ICPSR is a leading provider of social science data for research, offering a comprehensive archive of data sources. The link between the dataset and publication is manually curated, with a link established only when a publication significantly utilizes the datasets to produce results, as opposed to brief or tangential references. The initial dataset included 41,116 journal articles with DOI information available, with 37,440 published before 2020. Subsequently, we gather further information about each paper (e.g., citation counts, discipline, publication year, impact factor, references) and each author (e.g., author experience measured by number of citations) via the OpenAlex dataset in 2022~\cite{priem2022openalex}. Out of the 37,440  papers, 37,025 have records on OpenAlex and 30,479 papers have concept (subject) tags, references, and author information available. Ultimately, our final sample consists of 30,479 papers with 8,836 papers using multiple datasets and 21,643 papers only using one dataset, all with comprehensive information across all the aforementioned categories and published before 2020.
We also leverage the Altmetric dataset to identify mentions of research papers across multiple online sources, such as news, social media, policy documents, and Wikipedia. This dataset, provided by Altmetric (a full data dump as of August 31st, 2023), extensively monitors various online platforms to detect posts containing links or references to published research. Additional details can be found in the SI Appendix section 1.

\subsection*{Atypicality of dataset measurement}
To measure novelty, we adopt a general framework provided by the Rao-Stirling index~\cite{cassi2017analysing,Stirling2007-ya,Yang2022-zp}. Our novelty metric centers on dataset pairings within a paper, with infrequently paired datasets considered novel. From our dataset, we can calculate how often each pair of datasets has been used together in papers drawing from ICPSR data. Here, \emph{$Atypicality_c^{\text{Dataset}}$} represents the atypicality of dataset combinations of paper $c$. $d_c$ represents the set of datasets used in paper $c$ and $i, j \in d_c$. \emph{$D_{ij}$} represents the cosine similarity between dataset $i$ and $j$ in the data co-citation matrix. To compute \emph{$D_{ij}$}, we create an \emph{article vector} $h^a$ for each dataset $a$. Each coordinate in vector $h^a$ represents an article, and $h^a(b)$ is 1 if dataset $a$ is used in article $b$, and 0 otherwise. \emph{$D_{ij}$} is then defined as the cosine similarity between $h^i$ and $h^j$. Following the Rao-Stirling index, \emph{$P_i^c$} is the proportional representation of dataset $i$ in article $c$. In our case, we assume that all datasets contribute equally to the paper, and thus \emph{$P_i^c = \frac{1}{N_c}$}, where $N_c$ is the number of datasets in article $c$. $Atypicality_c^{\text{Dataset}}$ is defined in Equation~\ref{eq:novelty
}. In the regression analysis, we apply a z-score normalization to the atypicality score.

\begin{align}
Atypicality_c^{\text{Dataset}} &= 1-\sum_{ij \in d_c} D_{ij}*P_{i}^c*P_{j}^c \\
&= 1- \frac{1}{N_c^2} \sum_{ij \in d_c} D_{ij}
\label{eq:novelty}
\end{align}

\subsection*{Topic atypicality}


Adopting a similar quantification methodology as shown in Equation \ref{eq:novelty}, we first gather all topics covered by the datasets used in a single publication and calculate how often each pair of data topics has been used together amongst all ICPSR datasets.

Here, \emph{$Atypicality_c^{\text{Topic}}$} represents the topic atypicality of the dataset of paper $c$. $t_c$ represents the union of topics covered by the datasets used in paper $c$, and $u, v \in t_c$, with \emph{$D_{uv}$} representing the cosine similarity between dataset topics, and \emph{$P_u^c$} representing the proportion of topics in a dataset. To compute \emph{$D_{uv}$}, we create an \emph{article vector} $h^t$ for each topic $t$. Each coordinate of the vector $h^t$ represents each dataset's topic $m$ in our sample, where each coordinate corresponds to an article, and $h^t(m)$ is 1 if topic $t$ is used in article $m$, and 0 otherwise. With the article vectors defined, we can compare how different topics are utilized in the article. We calculate \emph{$D_{uv}$} by computing the cosine similarity between $h^u$ and $h^v$. \emph{$P_u^c$} is the proportional representation of topic $u$ in article $c$ (i.e., \emph{$P_u^c = \frac{1}{N_c}$}, where $N_c$ is the total number of topics in article $c$, assuming an equal contribution of each topic to article $c$). $Atypicality_c^{\text{Topic}}$ is defined in Equation~\ref{eq:novelty_2}.

\begin{align}
Atypicality_c^{\text{topic}} &= 1-\sum_{uv\in t_c} D_{uv}*P_{u}^c*P_{v}^c \\
&= 1- \frac{1}{N_c^2} \sum_{uv\in t_c} D_{uv}
\label{eq:novelty_2}
 \end{align}

 \subsection*{Paper novelty}

To measure the novelty of a paper, we use a similar approach to the one used to measure the atypicality of datasets (Equation \ref{eq:novelty}). Following prior work \cite{uzzi2013atypical}, our novelty measure centers on journal pairings referenced within a paper, with infrequently paired journals considered novel. We calculate how often each pair of journals has been referenced together in papers, drawing on OpenAlex data. 
\emph{$Atypicality_c^{\text{Journal}}$} represents the paper novelty of paper $c$.
$t_j$ represents the set of journals referenced in paper $c$, and $p, q \in t_j$. Here, \emph{$D_{pq}$} represents the cosine similarity between journals $p$ and $q$ in the journal co-citation matrix. To compute \emph{$D_{pq}$}, we create an \emph{article vector} $h^e$ for each journal $e$, with $h^e(b)$ being 1 if journal $e$ is referenced in article $b$, and 0 otherwise. With the article vectors defined, we can compare how different journals are utilized in the papers. We calculate \emph{$D_{pq}$} via the cosine similarity between $h^p$ and $h^q$. \emph{$R_{c}$} is the number of references in paper $c$, and \emph{$R_p^c$} is the number of references in paper $c$ that belong to journal $p$. \emph{$P_p^c$} is the proportional representation of journal $p$ in article $c$ (i.e., \emph{$P_p^c = \frac{R_p^c}{R_{c}}$}). $Atypicality_c^{\text{Journal}}$ is defined in Equation~\ref{eq:novelty2}
\begin{equation}
Atypicality_c^{\text{Journal}}= 1-\sum_{pq \in t_j} D_{pq}*P_{p}^c*P_{q}^c
\label{eq:novelty2}
 \end{equation}

\section*{Data, Materials, and Software Availability}
The Altmetric data can be accessed free of charge by researchers from \url{https://www.altmetric.com/research-access/}. The OpenAlex Dataset is publicly available at \url{https://docs.openalex.org/how-to-use-the-api/api-overview}  All custom code and data created by us have been deposited on GitHub (\url{https://github.com/yulin-yu/DataCombinations})

\section*{Acknowledgements} We are deeply grateful for the help and suggestions provided by Noshir Contractor, Christine Feak, Lizhou Fan, Libby Hemphill, Jane Im, Sara Lafia, Kerby Shedden, Kaicheng Yang, Jing Yang, Yian Yin, members of the Romero group, CSCAR, and the seminar and research talk participants from ICSSI 2023, the Equitable Opportunity Conference 2024, the Computational Social Science Working Group at the University of Michigan, Doctoral seminar (Special Topics in Information Science: Quantifying Scientific Ideas, Careers, and Teams) at Cornell University, and the Amaral Lab, the Lab on Innovation, Networks, and Knowledge, and Hyejin Youn's Lab at Northwestern University. We also extend our thanks to Altmetric.com for sharing the data used in this study.

\bibliography{cite}

\section*{Supplementary Information}

This document includes:

Supplementary Note 1: Data Description

Supplementary Note 2: Example of Measurements

Supplementary Note 3: Variable Description

Supplementary Note 4: Regression Tables

Supplementary Note 5: Regression Equations

Supplementary Note 6: Robustness Check - Matching
\newpage

\subsection*{1. Data Description}

We integrated three data resources to quantify the effect of the aytpicality of data combinations on scientific impact. 

(1) ICPSR Bibliography: a meticulous data source comprising social science datasets curated by ICPSR and publications published between 1962 and 2021 that cite these datasets. This link is established exclusively when the publication analyzes the data set or it includes a relevant discussion of data-related methodology~\cite{lafia2022subdivisions,hemphill2024dataset}. 

(2) OpenAlex dataset: a fully open scientific knowledge graph, encompassing metadata for 209 million works and 2,013 million authors. In this project, we relied on the OpenAlex dataset (OpenAlex API snapshot) to extract publication information, including references, citations, author lists, and disciplines. Additionally, we extracted author information, such as their total citations.

(3) Altmetric Dataset: The Altmetric Dataset captures online attention given to research publications. It encompasses approximately 191 million mentions of 35 million research outputs and identifies references to research papers from various online sources, including news articles, social media platforms, policy documents, and Wikipedia. We extracted Almetric mentions of all papers included in the analysis until August 2023. Since Altmetric started collecting data in 2011, we only extracted papers published after 2010. 

In total, we obtained 30,497 papers with 5,241 unique ICPSR datasets published before 2020.


\subsection*{2. Example of Measurements}

{\bf Atypicality of dataset combination:}

10 random examples from the top and bottom 25\% quantile of atypicality of data combination score:
\vspace{-\baselineskip} 
\includepdf[pages=-]{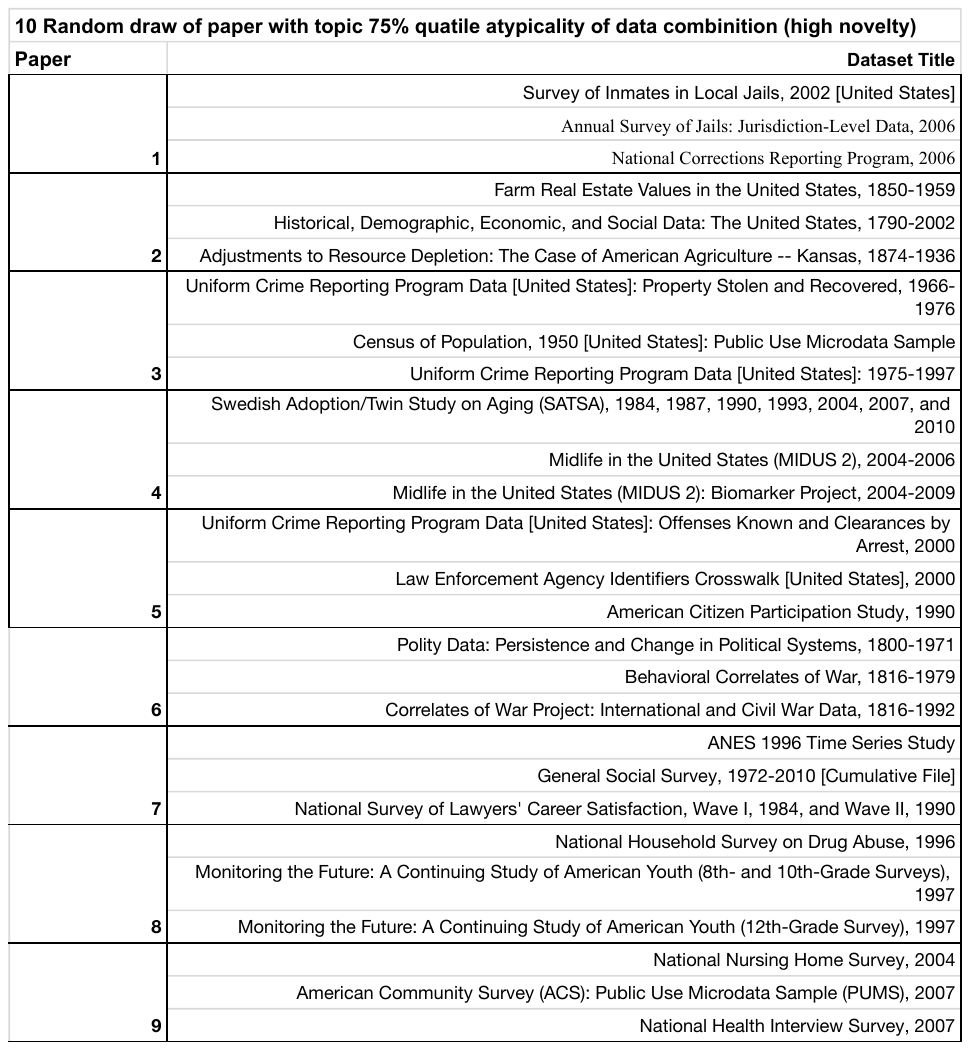}
\includepdf[pages=-]{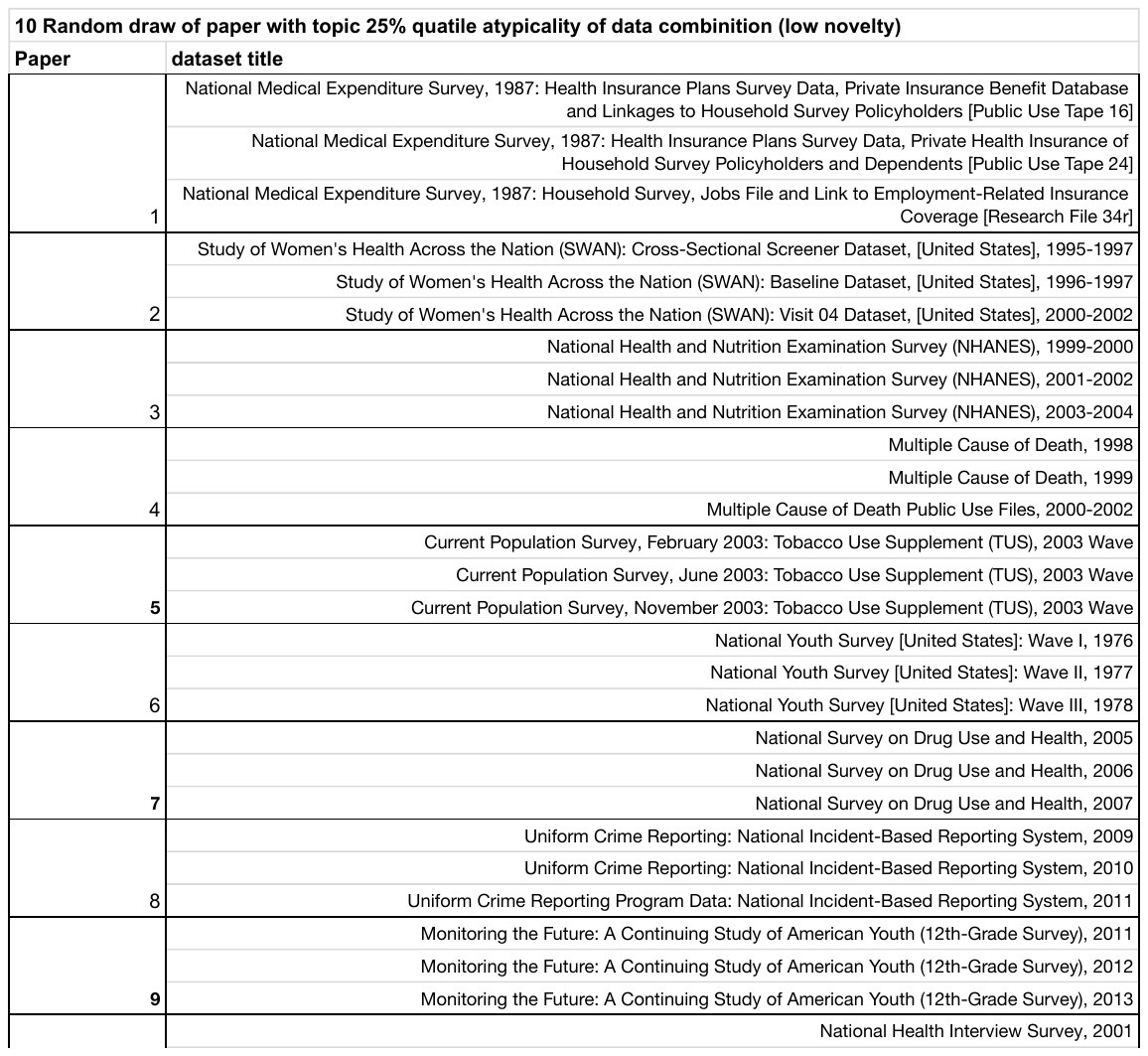}

{\bf Topic atypicality:}

10 ramdom examples of top and bottom 25\% quantile  topic atypicality quantile of topic atypicality
\includepdf[pages=-]{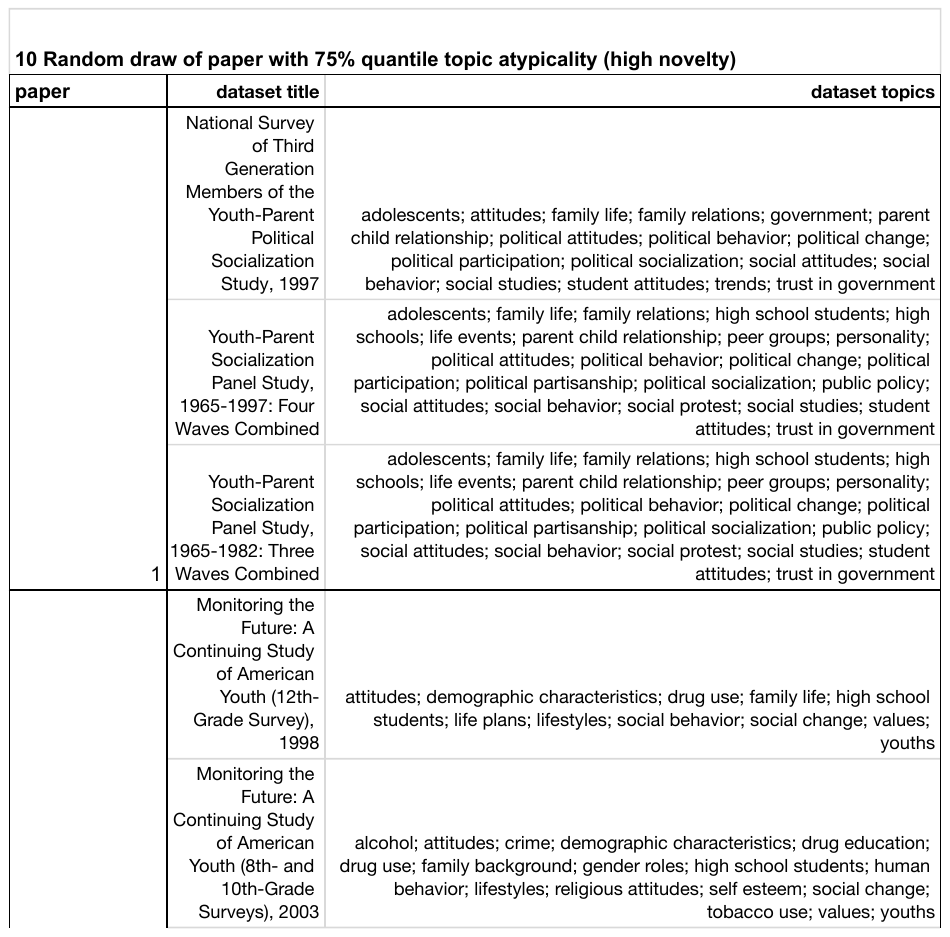}
\includepdf[pages=-]{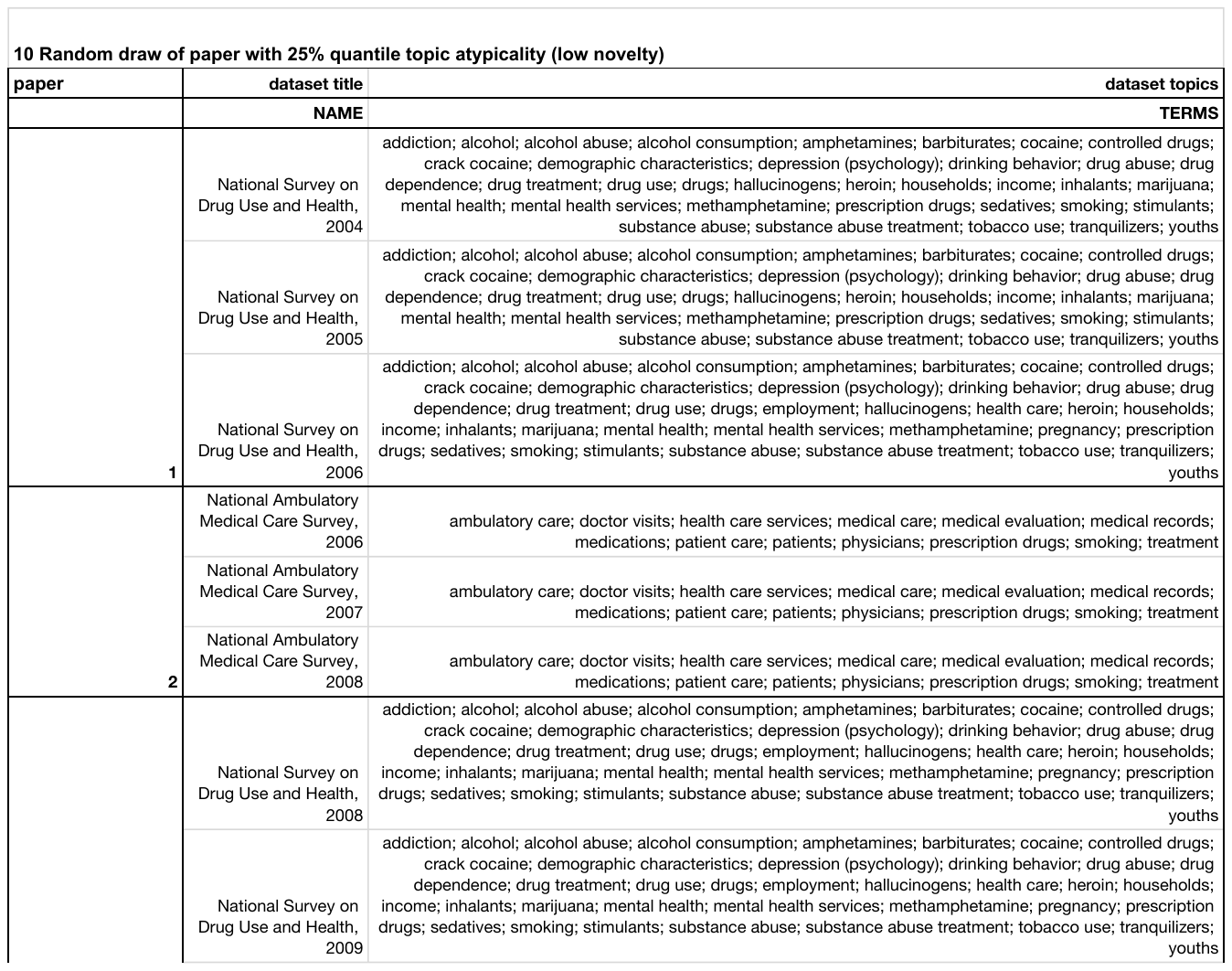}





\subsection*{3. Variable Descriptions}
\setcounter{figure}{0}

{\bf Number of datasets:} The total number of datasets used in a paper. The distribution of the number of datasets is shown in Figure S1.

\begin{figure}[htbp]
   \renewcommand{\figurename}{Figure S}
  \centering
  \includegraphics[width=0.8\textwidth]{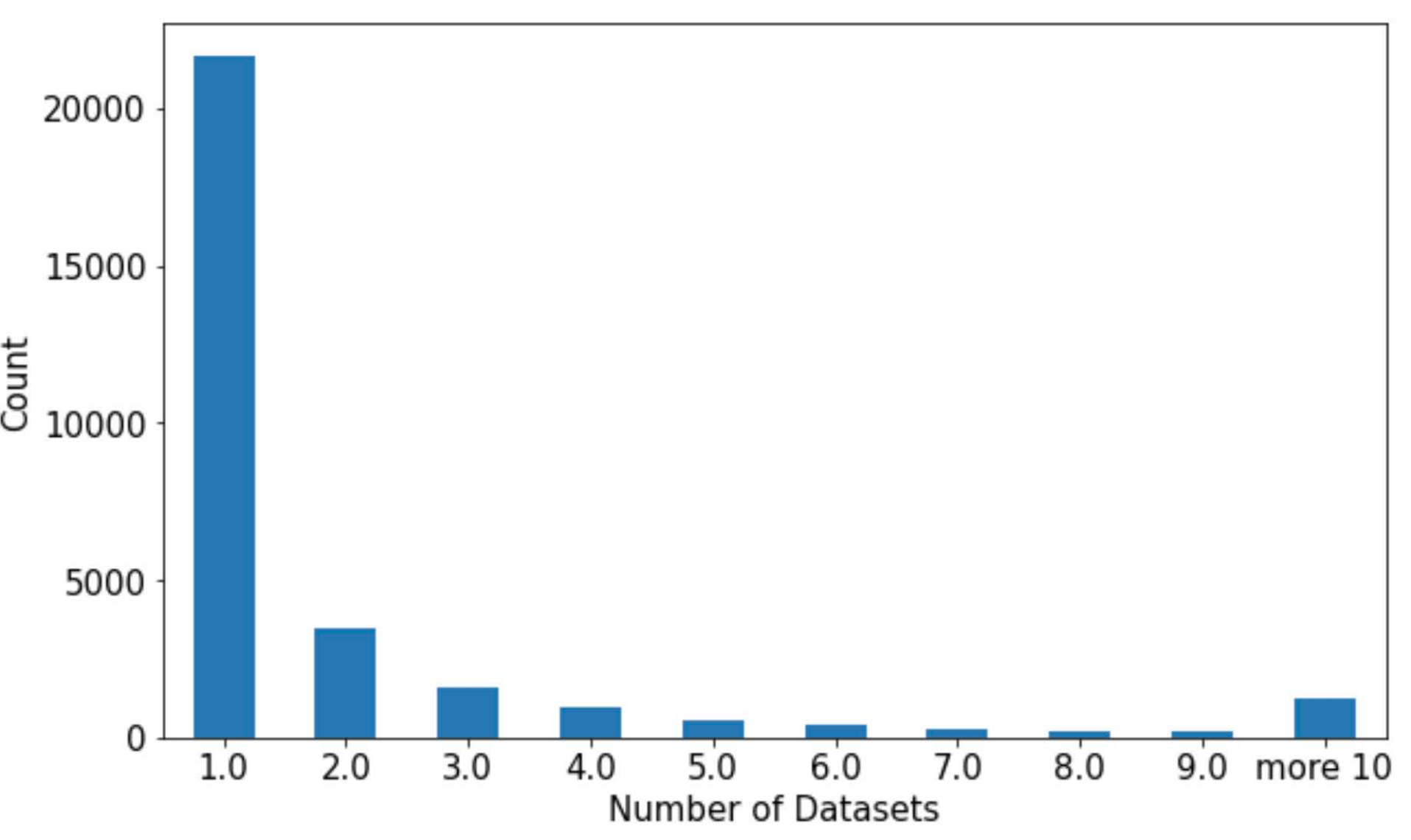}
  \caption{Distribution of number of datasets use}
  \label{fig:numd}
\end{figure}

\noindent {\bf 3,5,10 year citation counts:} We utilize the OpenAlex snapshot to extract all papers that have cited the targeted paper within a 3,5,10-year timeframe, starting from its publication year. The distribution of 3,5,10 year citations is shown in Figure S2.\\

\noindent {\bf 5\% hit paper:} We define a ``5\% hit paper" as a paper that has received citations within the top 5\% of all papers in our dataset, based on their citation count over a 3-year period.\\

\noindent {\bf Publication year:} The year of publication is significantly associated with citation rates. Therefore, in order to control for potential time effects, we incorporate the year variable as a dummy variable. Each dummy variable corresponds to a five-year interval, allowing us to effectively account for the impact of time on citation patterns. 
\begin{figure}[htbp]
   \renewcommand{\figurename}{Figure S}
  \centering
  \includegraphics[width=0.8\textwidth]{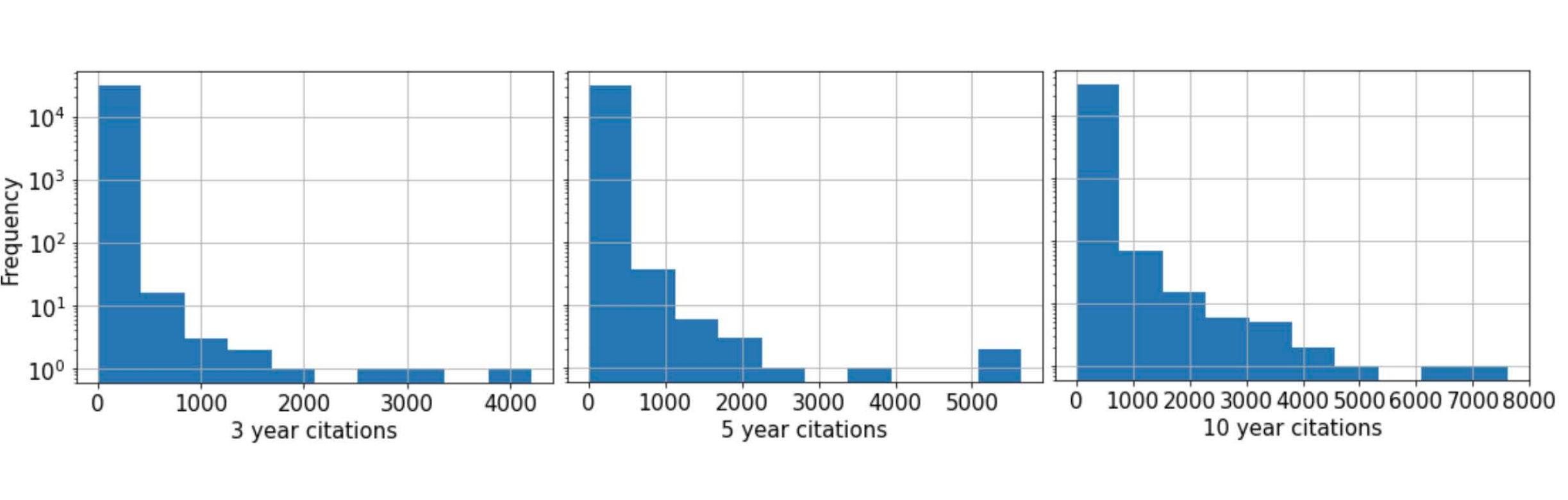}
  \caption{Distribution of 3, 5, and 10 year citation}
  \label{fig:labelname}
\end{figure}

\noindent {\bf Dataset use frequency:} A paper utilizing a frequently used dataset may focus on popular research questions, which could potentially confound the citation analysis. To address this concern, we introduce dataset use frequency as a controlling variable when investigating its impact on citation rates. In cases where a paper incorporates multiple datasets, we calculate the mean number of papers that utilize datasets used by a focal paper.\\

\noindent {\bf Number of authors:} Research on team science suggests that the number of co-authors positively correlates with citation impact~\cite{uzzi2013atypical,wu2019large}, as a larger number of co-authors tends to result in a more extensive citation network.\\

\noindent {\bf Author Recognition:} Author recognition can serve as a proxy for experience and authority in the field. Moreover, this variable has a strong correlation with citation impact. We measure author recognition of a paper as the average number of citations received by the authors. In the regression analysis, we log-transform the Author Recognition variable using Log(Author Recognition+0.01).\\

\noindent {\bf Disciplines:} We use the Level 0 disciplines provided by the OpenAlex dataset. Each paper in the dataset is associated with one or more discipline labels, and the weights for these labels are derived using deep learning models~\footnote{\url{https://docs.openalex.org/api-entities/concepts}}. The 19 major disciplines considered in the analysis are sociology, psychology, political science, physics, philosophy, medicine, mathematics, material science, history, geology, geography, environmental science, engineering, economics, computer science, chemistry, business, biology, and art. \\

\noindent {\bf Journal Impact Factor:} The majority of journals in our dataset do not have a publicly recorded impact factor. To address this limitation, we employ an alternative approach by calculating the average citation count for papers published in these journals in 2019. This average citation count is used as a proxy for the impact factor of the journal. In the regression analysis, we log-transform the Journal Impact Factor variable using Log(Journal Impact Factor+0.01).\\


\begin{figure}[htbp]
   \renewcommand{\figurename}{Figure S}
  \centering
  \includegraphics[width=0.8\textwidth]{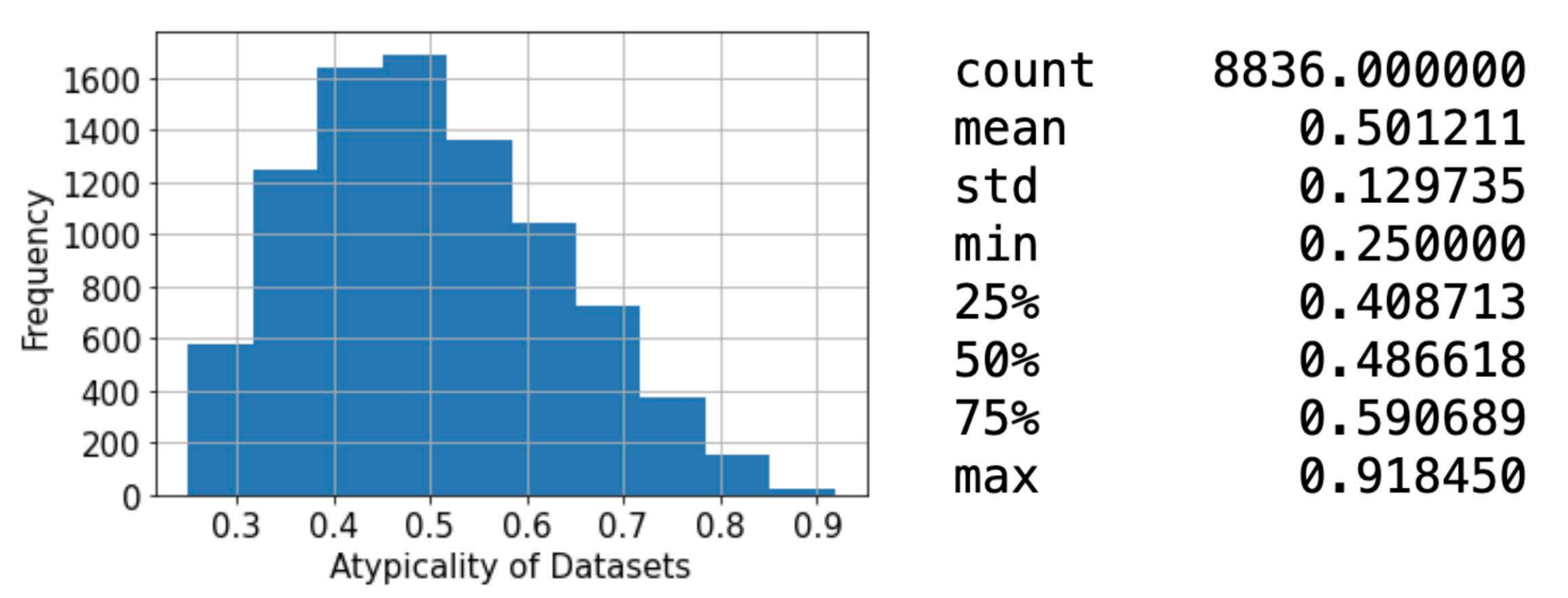}
  \caption{Distribution of atypicality of dataset}
  \label{fig:labelname}
\end{figure}


\subsection*{4. Regression Tables}

The full regression table is presented in Table 1 to Table 34.

\textbf{(1) The effect of dataset combinations on scientific impact: Impact of using multiple datasets (using data combinations) on citation over 3, 5, 10 years(Table 1-3).}

\FloatBarrier

\begin{table*}[ht]
\tiny
\begin{center}
  \renewcommand{\tablename}{Table S}

\begin{tabular}{lcccccc}
\textbf{Dep. Variable:}    3 year citation                                                            & \textbf{coef} & \textbf{std err} & \textbf{z} & \textbf{P$> |$z$|$} & \textbf{[0.025} & \textbf{0.975]}  \\
\hline
\hline
\textbf{Intercept}                      &       0.0534  &        0.099     &     0.541  &         0.588        &       -0.140    &        0.247     \\
\textbf{Year\_bin[T.(1974, 1979]]} &       0.1139  &        0.117     &     0.972  &         0.331        &       -0.116    &        0.343     \\
\textbf{Year\_bin[T.(1979, 1984]]} &       0.0403  &        0.102     &     0.394  &         0.694        &       -0.160    &        0.241     \\
\textbf{Year\_bin[T.(1984, 1989]]} &       0.1211  &        0.097     &     1.244  &         0.214        &       -0.070    &        0.312     \\
\textbf{Year\_bin[T.(1989, 1994]]} &       0.2466  &        0.095     &     2.597  &         0.009        &        0.061    &        0.433     \\
\textbf{Year\_bin[T.(1994, 1999]]} &       0.5664  &        0.093     &     6.106  &         0.000        &        0.385    &        0.748     \\
\textbf{Year\_bin[T.(1999, 2004]]} &       0.8551  &        0.092     &     9.292  &         0.000        &        0.675    &        1.035     \\
\textbf{Year\_bin[T.(2004, 2009]]} &       0.9675  &        0.092     &    10.544  &         0.000        &        0.788    &        1.147     \\
\textbf{Year\_bin[T.(2009, 2014]]} &       0.8408  &        0.092     &     9.177  &         0.000        &        0.661    &        1.020     \\
\textbf{Year\_bin[T.(2014, 2019]]} &       0.6245  &        0.092     &     6.806  &         0.000        &        0.445    &        0.804     \\
\textbf{binary\_UsingMultipleDataset}   &       0.1579  &        0.013     &    11.776  &         0.000        &        0.132    &        0.184     \\
\textbf{Data\_use\_frequency\_log}      &       0.0475  &        0.004     &    13.119  &         0.000        &        0.040    &        0.055     \\
\textbf{NumAuthor\_log}                 &       0.2926  &        0.011     &    26.753  &         0.000        &        0.271    &        0.314     \\
\textbf{AuthorExprience\_log}           &       0.0301  &        0.003     &     9.192  &         0.000        &        0.024    &        0.037     \\
\textbf{ImpactFactor\_log}              &       0.4152  &        0.010     &    43.360  &         0.000        &        0.396    &        0.434     \\
\textbf{Art}                            &      -1.7865  &        1.150     &    -1.554  &         0.120        &       -4.040    &        0.467     \\
\textbf{Biology}                        &       0.2856  &        0.173     &     1.647  &         0.100        &       -0.054    &        0.626     \\
\textbf{Business}                       &      -0.3956  &        0.075     &    -5.307  &         0.000        &       -0.542    &       -0.249     \\
\textbf{Chemistry}                      &      -0.0975  &        0.256     &    -0.382  &         0.703        &       -0.598    &        0.403     \\
\textbf{Computer\_science}              &       0.3527  &        0.106     &     3.313  &         0.001        &        0.144    &        0.561     \\
\textbf{Economics}                      &       0.3231  &        0.057     &     5.714  &         0.000        &        0.212    &        0.434     \\
\textbf{Engineering}                    &      -0.5778  &        0.179     &    -3.224  &         0.001        &       -0.929    &       -0.227     \\
\textbf{Environmental\_science}         &      -0.4148  &        0.296     &    -1.399  &         0.162        &       -0.996    &        0.166     \\
\textbf{Geography}                      &       0.0185  &        0.076     &     0.243  &         0.808        &       -0.130    &        0.167     \\
\textbf{Geology}                        &      -0.5313  &        3.343     &    -0.159  &         0.874        &       -7.084    &        6.021     \\
\textbf{History}                        &      -0.3322  &        0.245     &    -1.353  &         0.176        &       -0.813    &        0.149     \\
\textbf{Materials\_science}             &       0.7319  &        2.016     &     0.363  &         0.717        &       -3.219    &        4.683     \\
\textbf{Mathematics}                    &      -0.6005  &        0.150     &    -3.996  &         0.000        &       -0.895    &       -0.306     \\
\textbf{Medicine}                       &       0.5204  &        0.036     &    14.274  &         0.000        &        0.449    &        0.592     \\
\textbf{Philosophy}                     &      -0.2655  &        0.949     &    -0.280  &         0.780        &       -2.125    &        1.594     \\
\textbf{Physics}                        &      -0.1820  &        1.410     &    -0.129  &         0.897        &       -2.945    &        2.581     \\
\textbf{Political\_science}             &      -0.2279  &        0.053     &    -4.328  &         0.000        &       -0.331    &       -0.125     \\
\textbf{Psychology}                     &      -0.0128  &        0.040     &    -0.319  &         0.749        &       -0.091    &        0.066     \\
\textbf{Sociology}                      &       0.2663  &        0.055     &     4.878  &         0.000        &        0.159    &        0.373     \\
\hline

\textbf{  No. Observations:  } &     30479 & \textbf{  Log-Likelihood:    } &  -1.0842e+ & \textbf{  Dispersion:    } &  1  \\
\textbf{  Df Residuals:      } &     30445       & \textbf{  Df Model:          } &        33    \\
 \textbf{  Deviance:          } &  31325.  & \textbf{  Pearson chi2:      } &  7.49e+04
   \\

\end{tabular}
\caption{
Results of the negative binomial regression table with 3-Year Citations as the dependent variable and Using multiple datasets (1 indicates using more than one dataset in the publication, 0 otherwise) as the independent variable. } 
\end{center}
\end{table*}

\begin{table*}
\begin{center}
  \renewcommand{\tablename}{Table S}
\tiny 

\begin{tabular}{lcccccc}
\textbf{Dep. Variable:}    5 year citation                                                            & \textbf{coef} & \textbf{std err} & \textbf{z} & \textbf{P$> |$z$|$} & \textbf{[0.025} & \textbf{0.975]}  \\
\hline
\hline
\textbf{Intercept}                      &       0.6487  &        0.095     &     6.808  &         0.000        &        0.462    &        0.835     \\
\textbf{Year\_bin[T.(1974, 1979]]} &       0.1746  &        0.112     &     1.559  &         0.119        &       -0.045    &        0.394     \\
\textbf{Year\_bin[T.(1979, 1984]]} &       0.0677  &        0.098     &     0.691  &         0.489        &       -0.124    &        0.260     \\
\textbf{Year\_bin[T.(1984, 1989]]} &       0.1849  &        0.093     &     1.986  &         0.047        &        0.002    &        0.368     \\
\textbf{Year\_bin[T.(1989, 1994]]} &       0.3471  &        0.091     &     3.822  &         0.000        &        0.169    &        0.525     \\
\textbf{Year\_bin[T.(1994, 1999]]} &       0.7230  &        0.089     &     8.149  &         0.000        &        0.549    &        0.897     \\
\textbf{Year\_bin[T.(1999, 2004]]} &       0.9814  &        0.088     &    11.148  &         0.000        &        0.809    &        1.154     \\
\textbf{Year\_bin[T.(2004, 2009]]} &       1.0681  &        0.088     &    12.166  &         0.000        &        0.896    &        1.240     \\
\textbf{Year\_bin[T.(2009, 2014]]} &       0.8547  &        0.088     &     9.751  &         0.000        &        0.683    &        1.026     \\
\textbf{Year\_bin[T.(2014, 2019]]} &       0.6825  &        0.088     &     7.730  &         0.000        &        0.509    &        0.856     \\
\textbf{binary\_UsingMultipleDataset}   &       0.1415  &        0.014     &    10.211  &         0.000        &        0.114    &        0.169     \\
\textbf{Data\_use\_frequency\_log}      &       0.0495  &        0.004     &    13.427  &         0.000        &        0.042    &        0.057     \\
\textbf{NumAuthor\_log}                 &       0.2983  &        0.011     &    26.396  &         0.000        &        0.276    &        0.320     \\
\textbf{AuthorExprience\_log}           &       0.0257  &        0.003     &     7.685  &         0.000        &        0.019    &        0.032     \\
\textbf{ImpactFactor\_log}              &       0.3941  &        0.010     &    41.001  &         0.000        &        0.375    &        0.413     \\
\textbf{Art}                            &      -2.1903  &        1.055     &    -2.076  &         0.038        &       -4.258    &       -0.123     \\
\textbf{Biology}                        &      -0.0004  &        0.174     &    -0.002  &         0.998        &       -0.342    &        0.341     \\
\textbf{Business}                       &      -0.5457  &        0.076     &    -7.190  &         0.000        &       -0.694    &       -0.397     \\
\textbf{Chemistry}                      &      -0.1117  &        0.256     &    -0.436  &         0.663        &       -0.613    &        0.390     \\
\textbf{Computer\_science}              &       0.4511  &        0.109     &     4.126  &         0.000        &        0.237    &        0.665     \\
\textbf{Economics}                      &       0.3102  &        0.057     &     5.467  &         0.000        &        0.199    &        0.421     \\
\textbf{Engineering}                    &      -0.6599  &        0.177     &    -3.726  &         0.000        &       -1.007    &       -0.313     \\
\textbf{Environmental\_science}         &      -0.5090  &        0.297     &    -1.714  &         0.087        &       -1.091    &        0.073     \\
\textbf{Geography}                      &       0.0027  &        0.077     &     0.035  &         0.972        &       -0.149    &        0.154     \\
\textbf{Geology}                        &       0.3323  &        3.237     &     0.103  &         0.918        &       -6.012    &        6.677     \\
\textbf{History}                        &      -0.3825  &        0.245     &    -1.562  &         0.118        &       -0.862    &        0.097     \\
\textbf{Materials\_science}             &       1.1238  &        2.862     &     0.393  &         0.695        &       -4.485    &        6.733     \\
\textbf{Mathematics}                    &      -0.4460  &        0.153     &    -2.915  &         0.004        &       -0.746    &       -0.146     \\
\textbf{Medicine}                       &       0.4958  &        0.037     &    13.292  &         0.000        &        0.423    &        0.569     \\
\textbf{Philosophy}                     &       0.1352  &        0.886     &     0.153  &         0.879        &       -1.601    &        1.871     \\
\textbf{Physics}                        &      -1.0514  &        1.384     &    -0.760  &         0.447        &       -3.764    &        1.661     \\
\textbf{Political\_science}             &      -0.2567  &        0.053     &    -4.834  &         0.000        &       -0.361    &       -0.153     \\
\textbf{Psychology}                     &       0.0764  &        0.041     &     1.865  &         0.062        &       -0.004    &        0.157     \\
\textbf{Sociology}                      &       0.3766  &        0.055     &     6.863  &         0.000        &        0.269    &        0.484     \\
\hline

\textbf{  No. Observations:  } &     27833 & \textbf{  Log-Likelihood:    } & -1.1575e+05   & \textbf{  Dispersion:    } &  1\\
\textbf{  Df Residuals:      } &     27799       & \textbf{  Df Model:          } &        33    \\
 \textbf{  Deviance:          } &  28983    & \textbf{  Pearson chi2:      } &  7.00e+04 \\

\end{tabular}
\caption{
Results of the negative binomial  regression table with 5-Year Citations as the dependent variable and Using multiple datasets (1 indicates using more than one dataset in the publication, 0 otherwise) as independent variable. To capture 5-year citations, we track publications in our
dataset up to 2018 for this analysis.}  
\end{center}
\end{table*}

\begin{table*}
\begin{center}
  \renewcommand{\tablename}{Table S}
\tiny 

\begin{tabular}{lcccccc}
\textbf{Dep. Variable:}    10 year citation                                                            & \textbf{coef} & \textbf{std err} & \textbf{z} & \textbf{P$> |$z$|$} & \textbf{[0.025} & \textbf{0.975]}  \\
\hline
\hline
\textbf{Intercept}                      &       1.3986  &        0.095     &    14.732  &         0.000        &        1.212    &        1.585     \\
\textbf{Year\_bin[T.(1974, 1979]]} &       0.1540  &        0.109     &     1.418  &         0.156        &       -0.059    &        0.367     \\
\textbf{Year\_bin[T.(1979, 1984]]} &       0.0827  &        0.095     &     0.872  &         0.383        &       -0.103    &        0.268     \\
\textbf{Year\_bin[T.(1984, 1989]]} &       0.2895  &        0.090     &     3.210  &         0.001        &        0.113    &        0.466     \\
\textbf{Year\_bin[T.(1989, 1994]]} &       0.5693  &        0.088     &     6.473  &         0.000        &        0.397    &        0.742     \\
\textbf{Year\_bin[T.(1994, 1999]]} &       0.9385  &        0.086     &    10.917  &         0.000        &        0.770    &        1.107     \\
\textbf{Year\_bin[T.(1999, 2004]]} &       1.1891  &        0.085     &    13.936  &         0.000        &        1.022    &        1.356     \\
\textbf{Year\_bin[T.(2004, 2009]]} &       1.1448  &        0.085     &    13.443  &         0.000        &        0.978    &        1.312     \\
\textbf{Year\_bin[T.(2009, 2014]]} &       0.9075  &        0.086     &    10.602  &         0.000        &        0.740    &        1.075     \\
\textbf{binary\_UsingMultipleDataset}   &       0.1306  &        0.016     &     8.131  &         0.000        &        0.099    &        0.162     \\
\textbf{Data\_use\_frequency\_log}      &       0.0586  &        0.004     &    13.660  &         0.000        &        0.050    &        0.067     \\
\textbf{NumAuthor\_log}                 &       0.3211  &        0.013     &    24.185  &         0.000        &        0.295    &        0.347     \\
\textbf{AuthorExprience\_log}           &       0.0236  &        0.004     &     6.366  &         0.000        &        0.016    &        0.031     \\
\textbf{ImpactFactor\_log}              &       0.3556  &        0.011     &    33.244  &         0.000        &        0.335    &        0.377     \\
\textbf{Art}                            &      -4.8715  &        1.240     &    -3.927  &         0.000        &       -7.303    &       -2.440     \\
\textbf{Biology}                        &      -0.4305  &        0.203     &    -2.126  &         0.034        &       -0.827    &       -0.034     \\
\textbf{Business}                       &      -0.8555  &        0.086     &    -9.915  &         0.000        &       -1.025    &       -0.686     \\
\textbf{Chemistry}                      &      -0.1723  &        0.260     &    -0.662  &         0.508        &       -0.683    &        0.338     \\
\textbf{Computer\_science}              &       0.6471  &        0.128     &     5.069  &         0.000        &        0.397    &        0.897     \\
\textbf{Economics}                      &       0.3635  &        0.062     &     5.871  &         0.000        &        0.242    &        0.485     \\
\textbf{Engineering}                    &      -0.9793  &        0.192     &    -5.108  &         0.000        &       -1.355    &       -0.604     \\
\textbf{Environmental\_science}         &      -0.7360  &        0.334     &    -2.201  &         0.028        &       -1.391    &       -0.081     \\
\textbf{Geography}                      &      -0.0584  &        0.087     &    -0.669  &         0.504        &       -0.230    &        0.113     \\
\textbf{Geology}                        &       1.7228  &        3.195     &     0.539  &         0.590        &       -4.539    &        7.984     \\
\textbf{History}                        &      -0.5683  &        0.257     &    -2.216  &         0.027        &       -1.071    &       -0.066     \\
\textbf{Mathematics}                    &      -0.1561  &        0.180     &    -0.867  &         0.386        &       -0.509    &        0.197     \\
\textbf{Medicine}                       &       0.3670  &        0.043     &     8.619  &         0.000        &        0.284    &        0.451     \\
\textbf{Philosophy}                     &      -0.5310  &        0.858     &    -0.619  &         0.536        &       -2.213    &        1.151     \\
\textbf{Physics}                        &      -2.9997  &        1.605     &    -1.869  &         0.062        &       -6.146    &        0.147     \\
\textbf{Political\_science}             &      -0.3864  &        0.059     &    -6.596  &         0.000        &       -0.501    &       -0.272     \\
\textbf{Psychology}                     &       0.1743  &        0.047     &     3.698  &         0.000        &        0.082    &        0.267     \\
\textbf{Sociology}                      &       0.4898  &        0.060     &     8.108  &         0.000        &        0.371    &        0.608     \\
\hline

\textbf{  No. Observations:  } &     20401 & \textbf{  Log-Likelihood:    } &  -1.0141e+05 & \textbf{  Dispersion:    } &  1\\
\textbf{  Df Residuals:      } &    20369      & \textbf{  Df Model:          } &       31    \\
 \textbf{  Deviance:          } &  24092.    & \textbf{  Pearson chi2:      } & 6.03e+04 \\

\end{tabular}
\caption{
Results of the negative binomial regression table with 10-Year Citations as the dependent variable and Using multiple datasets (1 indicates using more than one dataset in the publication, 0 otherwise) as independent variable.  To capture 10-year citations, we track publications in our
dataset up to 2013 for this analysis.} 
\end{center}
\end{table*}

\FloatBarrier

\newpage
\textbf{Tables 4 - 7 present the effects of using multiple dataset (data combination)
on citations, based on a three-year analysis of publications released in four
distinct time periods: before 1990, 1990-2000, 2000-2010, and 2010-2020.}

\FloatBarrier

\begin{table*}[ht]
\begin{center}
  \renewcommand{\tablename}{Table S}
\tiny 

\begin{tabular}{lcccccc}
\textbf{Dep. Variable:}    3 year citation                                                            & \textbf{coef} & \textbf{std err} & \textbf{z} & \textbf{P$> |$z$|$} & \textbf{[0.025} & \textbf{0.975]}  \\
\hline
\hline
\textbf{Intercept}                    &       0.7340  &        0.066     &    11.184  &         0.000        &        0.605    &        0.863     \\
\textbf{binary\_UsingMultipleDataset} &       0.1823  &        0.019     &     9.594  &         0.000        &        0.145    &        0.220     \\
\textbf{Data\_use\_frequency\_log}    &       0.0336  &        0.005     &     6.537  &         0.000        &        0.024    &        0.044     \\
\textbf{NumAuthor\_log}               &       0.2394  &        0.015     &    15.662  &         0.000        &        0.209    &        0.269     \\
\textbf{AuthorExprience\_log}         &       0.0290  &        0.005     &     5.674  &         0.000        &        0.019    &        0.039     \\
\textbf{ImpactFactor\_log}            &       0.4825  &        0.015     &    33.026  &         0.000        &        0.454    &        0.511     \\
\textbf{Art}                          &       1.3826  &        2.182     &     0.634  &         0.526        &       -2.893    &        5.658     \\
\end{tabular}
\caption{
Results of the negative binomial regression table with 3-Year Citations as the dependent variable and Using multiple datasets (1 indicates using more than one dataset in the publication, 0 otherwise) as independent variable for paper published between 2010
and 2020.}
\end{center}
\end{table*}

\begin{table*}
\begin{center}
  \renewcommand{\tablename}{Table S}
\tiny 

\begin{tabular}{lcccccc}
\textbf{Dep. Variable:}    3 year citation                                                            & \textbf{coef} & \textbf{std err} & \textbf{z} & \textbf{P$> |$z$|$} & \textbf{[0.025} & \textbf{0.975]}  \\
\hline
\hline
\textbf{Intercept}                    &       0.9062  &        0.070     &    12.907  &         0.000        &        0.769    &        1.044     \\
\textbf{binary\_UsingMultipleDataset} &       0.1602  &        0.022     &     7.316  &         0.000        &        0.117    &        0.203     \\
\textbf{Data\_use\_frequency\_log}    &       0.0774  &        0.006     &    13.536  &         0.000        &        0.066    &        0.089     \\
\textbf{NumAuthor\_log}               &       0.3021  &        0.018     &    17.100  &         0.000        &        0.268    &        0.337     \\
\textbf{AuthorExprience\_log}         &       0.0270  &        0.005     &     5.116  &         0.000        &        0.017    &        0.037     \\
\textbf{ImpactFactor\_log}            &       0.4079  &        0.015     &    26.680  &         0.000        &        0.378    &        0.438     \\
\textbf{Art}                          &      -3.9904  &        2.090     &    -1.909  &         0.056        &       -8.087    &        0.106     \\
\end{tabular}
\caption{
Results of the negative binomial regression table with 3-Year Citations as the dependent variable and Using multiple datasets (1 indicates using more than one dataset in the publication, 0 otherwise) as independent variable for paper published between 2000
and 2010.}
\end{center}
\end{table*}

\begin{table*}[ht]
\begin{center}
  \renewcommand{\tablename}{Table S}
\tiny 

\begin{tabular}{lcccccc}
\textbf{Dep. Variable:}    3 year citation                                                            & \textbf{coef} & \textbf{std err} & \textbf{z} & \textbf{P$> |$z$|$} & \textbf{[0.025} & \textbf{0.975]}  \\
\hline
\hline
\textbf{Intercept}                    &       0.2679  &        0.102     &     2.623  &         0.009        &        0.068    &        0.468     \\
\textbf{binary\_UsingMultipleDataset} &       0.2560  &        0.035     &     7.288  &         0.000        &        0.187    &        0.325     \\
\textbf{Data\_use\_frequency\_log}    &       0.0802  &        0.011     &     7.599  &         0.000        &        0.060    &        0.101     \\
\textbf{NumAuthor\_log}               &       0.4601  &        0.030     &    15.310  &         0.000        &        0.401    &        0.519     \\
\textbf{AuthorExprience\_log}         &       0.0232  &        0.008     &     3.073  &         0.002        &        0.008    &        0.038     \\
\textbf{ImpactFactor\_log}            &       0.4449  &        0.023     &    19.216  &         0.000        &        0.399    &        0.490     \\
\textbf{Art}                          &     -47.5315  &     5.31e+04     &    -0.001  &         0.999        &    -1.04e+05    &     1.04e+05     \\
\end{tabular}
\caption{
Results of the negative binomial regression table with 3-Year Citations as the dependent variable and Using multiple datasets (1 indicates using more than one dataset in the publication, 0 otherwise) as independent variable for paper published between 1990
and 2000.}
\end{center}
\end{table*}

\begin{table*}
\begin{center}
  \renewcommand{\tablename}{Table S}
\tiny 

\begin{tabular}{lcccccc}
\textbf{Dep. Variable:}    3 year citation                                                            & \textbf{coef} & \textbf{std err} & \textbf{z} & \textbf{P$> |$z$|$} & \textbf{[0.025} & \textbf{0.975]}  \\
\hline
\hline
\textbf{Intercept}                    &       0.8658  &        0.145     &     5.991  &         0.000        &        0.583    &        1.149     \\
\textbf{binary\_UsingMultipleDataset} &       0.0073  &        0.056     &     0.131  &         0.896        &       -0.102    &        0.117     \\
\textbf{Data\_use\_frequency\_log}    &      -0.0285  &        0.016     &    -1.776  &         0.076        &       -0.060    &        0.003     \\
\textbf{NumAuthor\_log}               &       0.3980  &        0.050     &     7.909  &         0.000        &        0.299    &        0.497     \\
\textbf{AuthorExprience\_log}         &       0.0379  &        0.012     &     3.260  &         0.001        &        0.015    &        0.061     \\
\textbf{ImpactFactor\_log}            &       0.0819  &        0.035     &     2.362  &         0.018        &        0.014    &        0.150     \\
\textbf{Biology}                      &      -0.1719  &        0.889     &    -0.193  &         0.847        &       -1.914    &        1.570     \\
\end{tabular}
\caption{
Results of the negative binomial regression table with 3-Year Citations as the dependent variable and Using multiple datasets (1 indicates using more than one dataset in the publication, 0 otherwise) as independent variable for paper published before 1990.}
\end{center}
\end{table*}

\FloatBarrier
\textbf{Alternative Robustness check: Impact of number of datasets used on citation over 3, 5, 10 year. (Table 8-10)}
\FloatBarrier
\begin{table*}[ht]
\begin{center}
  \renewcommand{\tablename}{Table S}
\tiny 

\begin{tabular}{lcccccc}
  \textbf{Dep. Variable:} 3 year citation                                                              & \textbf{coef} & \textbf{std err} & \textbf{z} & \textbf{P$> |$z$|$} & \textbf{[0.025} & \textbf{0.975]}  \\
\hline
\hline
\textbf{Intercept}                      &       0.0329  &        0.099     &     0.333  &         0.739        &       -0.161    &        0.227     \\
\textbf{Year\_bin[T.(1974, 1979]]} &       0.1306  &        0.117     &     1.114  &         0.265        &       -0.099    &        0.360     \\
\textbf{Year\_bin[T.(1979, 1984]]} &       0.0662  &        0.102     &     0.647  &         0.518        &       -0.134    &        0.267     \\
\textbf{Year\_bin[T.(1984, 1989]]} &       0.1448  &        0.097     &     1.487  &         0.137        &       -0.046    &        0.336     \\
\textbf{Year\_bin[T.(1989, 1994]]} &       0.2673  &        0.095     &     2.815  &         0.005        &        0.081    &        0.453     \\
\textbf{Year\_bin[T.(1994, 1999]]} &       0.5869  &        0.093     &     6.326  &         0.000        &        0.405    &        0.769     \\
\textbf{Year\_bin[T.(1999, 2004]]} &       0.8673  &        0.092     &     9.421  &         0.000        &        0.687    &        1.048     \\
\end{tabular}
\caption{Results of the negative binomial regression table with 3-Year Citations as the dependent variable and number of dataset used as independent variable.  } 
\end{center}
\end{table*}

\begin{table*}
\begin{center}
  \renewcommand{\tablename}{Table S}
\tiny 

\begin{tabular}{lcccccc}
  \textbf{Dep. Variable:} 5 year citation                                                              & \textbf{coef} & \textbf{std err} & \textbf{z} & \textbf{P$> |$z$|$} & \textbf{[0.025} & \textbf{0.975]}  \\
\hline
\hline
\textbf{Intercept}                      &       0.6378  &        0.095     &     6.692  &         0.000        &        0.451    &        0.825     \\
\textbf{Year\_bin[T.(1974, 1979]]} &       0.1920  &        0.112     &     1.714  &         0.086        &       -0.028    &        0.411     \\
\textbf{Year\_bin[T.(1979, 1984]]} &       0.0926  &        0.098     &     0.947  &         0.344        &       -0.099    &        0.284     \\
\textbf{Year\_bin[T.(1984, 1989]]} &       0.2068  &        0.093     &     2.220  &         0.026        &        0.024    &        0.389     \\
\textbf{Year\_bin[T.(1989, 1994]]} &       0.3665  &        0.091     &     4.036  &         0.000        &        0.189    &        0.544     \\
\textbf{Year\_bin[T.(1994, 1999]]} &       0.7450  &        0.089     &     8.398  &         0.000        &        0.571    &        0.919     \\
\textbf{Year\_bin[T.(1999, 2004]]} &       0.9941  &        0.088     &    11.291  &         0.000        &        0.822    &        1.167     \\
\end{tabular}
\caption{Results of the negative binomial regression table with 5-Year Citations as the dependent variable and number of dataset used as independent variable.  To capture 5-year citations, we track publications in our
dataset up to 2018 for this analysis. }  
\end{center}
\end{table*}

\begin{table*}
\begin{center}
  \renewcommand{\tablename}{Table S}
\tiny 

\begin{tabular}{lcccccc}
  \textbf{Dep. Variable:} 10 year citation                                                              & \textbf{coef} & \textbf{std err} & \textbf{z} & \textbf{P$> |$z$|$} & \textbf{[0.025} & \textbf{0.975]}  \\
\hline
\hline
\textbf{Intercept}                      &       1.3949  &        0.095     &    14.691  &         0.000        &        1.209    &        1.581     \\
\textbf{Year\_bin[T.(1974, 1979]]} &       0.1713  &        0.109     &     1.578  &         0.115        &       -0.041    &        0.384     \\
\textbf{Year\_bin[T.(1979, 1984]]} &       0.1081  &        0.095     &     1.141  &         0.254        &       -0.078    &        0.294     \\
\textbf{Year\_bin[T.(1984, 1989]]} &       0.3108  &        0.090     &     3.448  &         0.001        &        0.134    &        0.487     \\
\textbf{Year\_bin[T.(1989, 1994]]} &       0.5896  &        0.088     &     6.707  &         0.000        &        0.417    &        0.762     \\
\textbf{Year\_bin[T.(1994, 1999]]} &       0.9636  &        0.086     &    11.215  &         0.000        &        0.795    &        1.132     \\
\textbf{Year\_bin[T.(1999, 2004]]} &       1.2052  &        0.085     &    14.128  &         0.000        &        1.038    &        1.372     \\
\end{tabular}
\caption{Results of the negative binomial table with 10-Year Citations as the dependent variable and number of dataset used as independent variable.  To capture 10-year citations, we track publications in our
dataset up to 2013 for this analysis. } 
\end{center}
\end{table*}

\FloatBarrier

\FloatBarrier
\textbf{(2) Atypical combinations of datasets associate with high impact: - Impact of using atypical combinations of datasets on citation over 3, 5, 10 year. (Table 11-13)}

\FloatBarrier

\begin{table*}[ht]
\begin{center}
  \renewcommand{\tablename}{Table S} 

\tiny 

\begin{tabular}{lcccccc}
\tiny 
  \textbf{Dep. Variable:}3 year citation & \textbf{coef} & \textbf{std err} & \textbf{z} & \textbf{P$> |$z$|$} & \textbf{[0.025} & \textbf{0.975]}  \\
\hline
\hline
\textbf{Intercept}                      &       0.0992  &        0.313     &     0.317  &         0.751        &       -0.515    &        0.713     \\
\textbf{Year\_bin[T.(1974, 1979]]} &       0.0454  &        0.339     &     0.134  &         0.893        &       -0.619    &        0.709     \\
\textbf{Year\_bin[T.(1979, 1984]]} &       0.2341  &        0.315     &     0.744  &         0.457        &       -0.383    &        0.851     \\
\textbf{Year\_bin[T.(1984, 1989]]} &       0.2221  &        0.310     &     0.717  &         0.473        &       -0.385    &        0.829     \\
\textbf{Year\_bin[T.(1989, 1994]]} &       0.4032  &        0.308     &     1.310  &         0.190        &       -0.200    &        1.007     \\
\textbf{Year\_bin[T.(1994, 1999]]} &       0.7818  &        0.305     &     2.561  &         0.010        &        0.184    &        1.380     \\
\textbf{Year\_bin[T.(1999, 2004]]} &       0.8743  &        0.305     &     2.868  &         0.004        &        0.277    &        1.472     \\
\textbf{Year\_bin[T.(2004, 2009]]} &       1.0178  &        0.304     &     3.343  &         0.001        &        0.421    &        1.614     \\
\textbf{Year\_bin[T.(2009, 2014]]} &       0.8967  &        0.304     &     2.947  &         0.003        &        0.300    &        1.493     \\
\textbf{Year\_bin[T.(2014, 2019]]} &       0.6933  &        0.304     &     2.278  &         0.023        &        0.097    &        1.290     \\
\textbf{Atypicality\_of\_datasets}      &       0.1692  &        0.018     &     9.569  &         0.000        &        0.135    &        0.204     \\
\textbf{NumDatasets\_log}               &      -0.1002  &        0.024     &    -4.159  &         0.000        &       -0.147    &       -0.053     \\
\textbf{Paper\_novelty}                 &       0.1368  &        0.013     &    10.650  &         0.000        &        0.112    &        0.162     \\
\textbf{Data\_use\_frequency\_log}      &       0.0791  &        0.009     &     8.338  &         0.000        &        0.061    &        0.098     \\
\textbf{NumAuthor\_log}                 &       0.3255  &        0.021     &    15.394  &         0.000        &        0.284    &        0.367     \\
\textbf{AuthorExprience\_log}           &       0.0098  &        0.006     &     1.607  &         0.108        &       -0.002    &        0.022     \\
\textbf{ImpactFactor\_log}              &       0.4086  &        0.018     &    23.174  &         0.000        &        0.374    &        0.443     \\
\textbf{Biology}                        &      -0.0339  &        0.473     &    -0.072  &         0.943        &       -0.962    &        0.894     \\
\textbf{Business}                       &       0.2118  &        0.135     &     1.572  &         0.116        &       -0.052    &        0.476     \\
\textbf{Chemistry}                      &       0.7228  &        0.433     &     1.671  &         0.095        &       -0.125    &        1.571     \\
\textbf{Computer\_science}              &       0.1888  &        0.219     &     0.862  &         0.389        &       -0.240    &        0.618     \\
\textbf{Economics}                      &       0.5756  &        0.099     &     5.825  &         0.000        &        0.382    &        0.769     \\
\textbf{Engineering}                    &      -1.0857  &        0.307     &    -3.537  &         0.000        &       -1.687    &       -0.484     \\
\textbf{Environmental\_science}         &       0.2907  &        0.459     &     0.634  &         0.526        &       -0.609    &        1.190     \\
\textbf{Geography}                      &       0.3730  &        0.132     &     2.825  &         0.005        &        0.114    &        0.632     \\
\textbf{History}                        &       0.0028  &        0.407     &     0.007  &         0.995        &       -0.796    &        0.801     \\
\textbf{Materials\_science}             &      -0.2305  &        2.944     &    -0.078  &         0.938        &       -6.000    &        5.539     \\
\textbf{Mathematics}                    &      -0.8843  &        0.333     &    -2.659  &         0.008        &       -1.536    &       -0.232     \\
\textbf{Medicine}                       &       0.8148  &        0.067     &    12.149  &         0.000        &        0.683    &        0.946     \\
\textbf{Philosophy}                     &      -0.1637  &        2.010     &    -0.081  &         0.935        &       -4.104    &        3.777     \\
\textbf{Political\_science}             &      -0.0888  &        0.093     &    -0.958  &         0.338        &       -0.271    &        0.093     \\
\textbf{Psychology}                     &       0.1479  &        0.075     &     1.984  &         0.047        &        0.002    &        0.294     \\
\textbf{Sociology}                      &       0.4538  &        0.107     &     4.229  &         0.000        &        0.243    &        0.664     \\
\hline
\textbf{  No. Observations:  } &     8836   & \textbf{  Log-Likelihood:    } &    -32637.& \textbf{  Dispersion:    } &  1 \\
 \textbf{  Df Residuals:      } &     8803      & \textbf{  Df Model:          } &       32    \\
\textbf{  Pearson chi2:      } &   2.46e+04& \textbf{  Deviance:          } &     9462.3  \\

\end{tabular}
\caption{
Results of the negative binomial regression table with 3-Year Citations as the dependent variable and Atypicality of Data Combinations as independent variable. }  
\end{center}
\end{table*}

\begin{table*}
\begin{center}

  \renewcommand{\tablename}{Table S}
\tiny 

\begin{tabular}{lcccccc}
\tiny 
                                                  \textbf{Dep. Variable:}  5 year citation             & \textbf{coef} & \textbf{std err} & \textbf{z} & \textbf{P$> |$z$|$} & \textbf{[0.025} & \textbf{0.975]}  \\
\hline
\hline
\textbf{Intercept}                      &       0.9925  &        0.292     &     3.396  &         0.001        &        0.420    &        1.565     \\
\textbf{Year\_bin[T.(1974, 1979]]} &       0.1346  &        0.315     &     0.428  &         0.669        &       -0.482    &        0.751     \\
\textbf{Year\_bin[T.(1979, 1984]]} &       0.1369  &        0.292     &     0.468  &         0.640        &       -0.436    &        0.710     \\
\textbf{Year\_bin[T.(1984, 1989]]} &       0.1398  &        0.287     &     0.487  &         0.626        &       -0.423    &        0.703     \\
\textbf{Year\_bin[T.(1989, 1994]]} &       0.3520  &        0.286     &     1.233  &         0.218        &       -0.208    &        0.912     \\
\textbf{Year\_bin[T.(1994, 1999]]} &       0.8444  &        0.283     &     2.984  &         0.003        &        0.290    &        1.399     \\
\textbf{Year\_bin[T.(1999, 2004]]} &       0.8755  &        0.283     &     3.098  &         0.002        &        0.322    &        1.429     \\
\textbf{Year\_bin[T.(2004, 2009]]} &       0.9798  &        0.282     &     3.471  &         0.001        &        0.427    &        1.533     \\
\textbf{Year\_bin[T.(2009, 2014]]} &       0.7903  &        0.282     &     2.802  &         0.005        &        0.237    &        1.343     \\
\textbf{Year\_bin[T.(2014, 2019]]} &       0.6120  &        0.283     &     2.165  &         0.030        &        0.058    &        1.166     \\
\textbf{Atypicality\_of\_datasets}      &       0.1747  &        0.018     &     9.579  &         0.000        &        0.139    &        0.210     \\
\textbf{NumDatasets\_log}               &      -0.1389  &        0.025     &    -5.543  &         0.000        &       -0.188    &       -0.090     \\
\textbf{Paper\_novelty}                 &       0.1627  &        0.013     &    12.493  &         0.000        &        0.137    &        0.188     \\
\textbf{Data\_use\_frequency\_log}      &       0.0679  &        0.010     &     6.929  &         0.000        &        0.049    &        0.087     \\
\textbf{NumAuthor\_log}                 &       0.3284  &        0.022     &    15.013  &         0.000        &        0.286    &        0.371     \\
\textbf{AuthorExprience\_log}           &      -0.0035  &        0.006     &    -0.560  &         0.575        &       -0.016    &        0.009     \\
\textbf{ImpactFactor\_log}              &       0.3912  &        0.018     &    21.889  &         0.000        &        0.356    &        0.426     \\
\textbf{Biology}                        &       0.2526  &        0.473     &     0.534  &         0.594        &       -0.675    &        1.180     \\
\textbf{Business}                       &       0.0735  &        0.141     &     0.522  &         0.602        &       -0.202    &        0.349     \\
\textbf{Chemistry}                      &       0.7253  &        0.436     &     1.664  &         0.096        &       -0.129    &        1.579     \\
\textbf{Computer\_science}              &       0.3542  &        0.226     &     1.565  &         0.118        &       -0.089    &        0.798     \\
\textbf{Economics}                      &       0.4671  &        0.100     &     4.652  &         0.000        &        0.270    &        0.664     \\
\textbf{Engineering}                    &      -1.0480  &        0.304     &    -3.446  &         0.001        &       -1.644    &       -0.452     \\
\textbf{Environmental\_science}         &       0.2787  &        0.458     &     0.609  &         0.543        &       -0.618    &        1.176     \\
\textbf{Geography}                      &       0.2969  &        0.135     &     2.205  &         0.027        &        0.033    &        0.561     \\
\textbf{History}                        &       0.0198  &        0.419     &     0.047  &         0.962        &       -0.801    &        0.841     \\
\textbf{Materials\_science}             &       1.1040  &        2.897     &     0.381  &         0.703        &       -4.574    &        6.782     \\
\textbf{Mathematics}                    &      -0.7677  &        0.328     &    -2.338  &         0.019        &       -1.411    &       -0.124     \\
\textbf{Medicine}                       &       0.8057  &        0.069     &    11.646  &         0.000        &        0.670    &        0.941     \\
\textbf{Philosophy}                     &       1.8757  &        1.687     &     1.112  &         0.266        &       -1.432    &        5.183     \\
\textbf{Political\_science}             &      -0.1519  &        0.095     &    -1.607  &         0.108        &       -0.337    &        0.033     \\
\textbf{Psychology}                     &       0.1923  &        0.077     &     2.491  &         0.013        &        0.041    &        0.344     \\
\textbf{Sociology}                      &       0.5774  &        0.109     &     5.299  &         0.000        &        0.364    &        0.791     \\
\hline
\textbf{  No. Observations:  } &     7975  & \textbf{  Log-Likelihood:    } &  -34101. & \textbf{  Dispersion:    } &  1  \\
\textbf{  Df Model:          } &       32     & \textbf{  Df Residuals:      } &     7942   \\
 \textbf{  Pearson chi2:      } &   2.07e+04    & \textbf{  Deviance:          } &    8493.0 \\

\end{tabular}
\caption{the Negative Binomial Regression Model Detail: the Effects of Atypicality of Data Combinations on 5-Year Citations.  {\it Note:} This study serves as a robustness test. To capture 5-year citations, we track publications in our dataset up to 2018 for this analysis. }   
\end{center}
\end{table*}

\begin{table*}
\begin{center}
  \renewcommand{\tablename}{Table S}
\tiny 

\begin{tabular}{lcccccc}
\tiny 
                                                  \textbf{Dep. Variable:}  10 year citation             & \textbf{coef} & \textbf{std err} & \textbf{z} & \textbf{P$> |$z$|$} & \textbf{[0.025} & \textbf{0.975]}  \\
\hline
\hline
\textbf{Intercept}                      &       1.9833  &        0.283     &     7.018  &         0.000        &        1.429    &        2.537     \\
\textbf{Year\_bin[T.(1974, 1979]]} &       0.0191  &        0.300     &     0.063  &         0.949        &       -0.570    &        0.608     \\
\textbf{Year\_bin[T.(1979, 1984]]} &       0.0124  &        0.279     &     0.044  &         0.965        &       -0.534    &        0.559     \\
\textbf{Year\_bin[T.(1984, 1989]]} &       0.0668  &        0.274     &     0.244  &         0.807        &       -0.469    &        0.603     \\
\textbf{Year\_bin[T.(1989, 1994]]} &       0.4237  &        0.272     &     1.557  &         0.119        &       -0.109    &        0.957     \\
\textbf{Year\_bin[T.(1994, 1999]]} &       0.9495  &        0.270     &     3.522  &         0.000        &        0.421    &        1.478     \\
\textbf{Year\_bin[T.(1999, 2004]]} &       0.9711  &        0.269     &     3.606  &         0.000        &        0.443    &        1.499     \\
\textbf{Year\_bin[T.(2004, 2009]]} &       0.9295  &        0.269     &     3.454  &         0.001        &        0.402    &        1.457     \\
\textbf{Year\_bin[T.(2009, 2014]]} &       0.7469  &        0.269     &     2.772  &         0.006        &        0.219    &        1.275     \\
\textbf{Atypicality\_of\_datasets}      &       0.1811  &        0.021     &     8.695  &         0.000        &        0.140    &        0.222     \\
\textbf{NumDatasets\_log}               &      -0.1756  &        0.029     &    -5.966  &         0.000        &       -0.233    &       -0.118     \\
\textbf{Paper\_novelty}                 &       0.1801  &        0.014     &    12.513  &         0.000        &        0.152    &        0.208     \\
\textbf{Data\_use\_frequency\_log}      &       0.0474  &        0.011     &     4.237  &         0.000        &        0.025    &        0.069     \\
\textbf{NumAuthor\_log}                 &       0.3539  &        0.026     &    13.766  &         0.000        &        0.303    &        0.404     \\
\textbf{AuthorExprience\_log}           &      -0.0028  &        0.007     &    -0.394  &         0.694        &       -0.017    &        0.011     \\
\textbf{ImpactFactor\_log}              &       0.3515  &        0.020     &    17.776  &         0.000        &        0.313    &        0.390     \\
\textbf{Biology}                        &       0.4735  &        0.553     &     0.856  &         0.392        &       -0.610    &        1.557     \\
\textbf{Business}                       &      -0.2941  &        0.159     &    -1.853  &         0.064        &       -0.605    &        0.017     \\
\textbf{Chemistry}                      &       0.8520  &        0.458     &     1.861  &         0.063        &       -0.045    &        1.750     \\
\textbf{Computer\_science}              &      -0.1202  &        0.281     &    -0.428  &         0.669        &       -0.670    &        0.430     \\
\textbf{Economics}                      &       0.5399  &        0.110     &     4.890  &         0.000        &        0.324    &        0.756     \\
\textbf{Engineering}                    &      -1.4427  &        0.324     &    -4.457  &         0.000        &       -2.077    &       -0.808     \\
\textbf{Environmental\_science}         &       0.0651  &        0.538     &     0.121  &         0.904        &       -0.989    &        1.119     \\
\textbf{Geography}                      &       0.1940  &        0.154     &     1.260  &         0.208        &       -0.108    &        0.496     \\
\textbf{History}                        &      -0.6619  &        0.464     &    -1.428  &         0.153        &       -1.571    &        0.247     \\
\textbf{Mathematics}                    &      -0.5465  &        0.355     &    -1.539  &         0.124        &       -1.242    &        0.149     \\
\textbf{Medicine}                       &       0.7159  &        0.079     &     9.015  &         0.000        &        0.560    &        0.872     \\
\textbf{Philosophy}                     &       0.3825  &        1.642     &     0.233  &         0.816        &       -2.836    &        3.601     \\
\textbf{Political\_science}             &      -0.3297  &        0.105     &    -3.153  &         0.002        &       -0.535    &       -0.125     \\
\textbf{Psychology}                     &       0.2735  &        0.090     &     3.024  &         0.002        &        0.096    &        0.451     \\
\textbf{Sociology}                      &       0.8238  &        0.122     &     6.759  &         0.000        &        0.585    &        1.063     \\
\hline
\textbf{  No. Observations:  } &     5815  & \textbf{  Log-Likelihood:    } &   -29348.  & \textbf{  Dispersion:    } &  1  \\
\textbf{  Df Model:          } &       30     & \textbf{  Df Residuals:      } &    5784    \\
 \textbf{  Pearson chi2:      } &    1.58e+04    & \textbf{  Deviance:          } &   6836.6   \\

\end{tabular}
\caption{the Negative Binomial Regression Model Detail: the Effects of Atypicality of Data Combinations on 10-Year Citations.  {\it Note:} This study serves as a robustness test. To capture 10-year citations, we track publications in our dataset up to 2013 for this analysis. }   
\end{center}
\end{table*}

\FloatBarrier

\textbf{- Tables 14-17 present the effects of using atypical combinations of datasets on citations, based on a three-year analysis of publications released in four distinct time periods: before 1990, 1990-2000, 2000-2010, and 2010-2020.}

\FloatBarrier

\begin{table*}[ht]
\begin{center}
  \renewcommand{\tablename}{Table S}
\tiny 

\begin{tabular}{lcccccc}
\tiny 
                                                  \textbf{Dep. Variable:}  3 year citation             & \textbf{coef} & \textbf{std err} & \textbf{z} & \textbf{P$> |$z$|$} & \textbf{[0.025} & \textbf{0.975]}  \\
\hline
\hline
\textbf{Intercept}                 &       0.6864  &        0.131     &     5.254  &         0.000        &        0.430    &        0.942     \\
\textbf{Atypicality\_of\_datasets} &       0.2003  &        0.026     &     7.831  &         0.000        &        0.150    &        0.250     \\
\textbf{NumDatasets\_log}          &      -0.0914  &        0.034     &    -2.723  &         0.006        &       -0.157    &       -0.026     \\
\textbf{Paper\_novelty}            &       0.1450  &        0.020     &     7.142  &         0.000        &        0.105    &        0.185     \\
\textbf{Data\_use\_frequency\_log} &       0.0938  &        0.014     &     6.917  &         0.000        &        0.067    &        0.120     \\
\textbf{NumAuthor\_log}            &       0.3006  &        0.029     &    10.409  &         0.000        &        0.244    &        0.357     \\
\textbf{AuthorExprience\_log}      &       0.0060  &        0.009     &     0.667  &         0.505        &       -0.012    &        0.024     \\
\end{tabular}
\caption{the Negative Binomial Regression Model Detail: the Effects of Atypicality of Data Combinations on 3-Year Citations for paper published between 2010 to 2020. }
\end{center}
\end{table*}

\begin{table*}
\begin{center}

  \renewcommand{\tablename}{Table S}
\tiny 

\begin{tabular}{lcccccc}
\tiny 
                                                  \textbf{Dep. Variable:}  3 year citation             & \textbf{coef} & \textbf{std err} & \textbf{z} & \textbf{P$> |$z$|$} & \textbf{[0.025} & \textbf{0.975]}  \\
\hline
\hline
\textbf{Intercept}                 &       1.1344  &        0.145     &     7.838  &         0.000        &        0.851    &        1.418     \\
\textbf{Atypicality\_of\_datasets} &       0.1929  &        0.028     &     6.778  &         0.000        &        0.137    &        0.249     \\
\textbf{NumDatasets\_log}          &      -0.1589  &        0.041     &    -3.912  &         0.000        &       -0.238    &       -0.079     \\
\textbf{Paper\_novelty}            &       0.1706  &        0.021     &     7.991  &         0.000        &        0.129    &        0.212     \\
\textbf{Data\_use\_frequency\_log} &       0.0846  &        0.015     &     5.606  &         0.000        &        0.055    &        0.114     \\
\textbf{NumAuthor\_log}            &       0.2898  &        0.034     &     8.418  &         0.000        &        0.222    &        0.357     \\
\textbf{AuthorExprience\_log}      &       0.0071  &        0.010     &     0.693  &         0.488        &       -0.013    &        0.027     \\
\end{tabular}
\caption{the Negative Binomial Regression Model Detail: the Effects of Atypicality of Data Combinations on 3-Year Citations for paper published between 2000 to 2010. }
    
\end{center}
\end{table*}

\begin{table*}
  \renewcommand{\tablename}{Table S}
\tiny 
\begin{center}
\begin{tabular}{lcccccc}
\tiny 
                                                  \textbf{Dep. Variable:}  3 year citation             & \textbf{coef} & \textbf{std err} & \textbf{z} & \textbf{P$> |$z$|$} & \textbf{[0.025} & \textbf{0.975]}  \\
\hline
\hline
\textbf{Intercept}                 &       0.7209  &        0.212     &     3.394  &         0.001        &        0.305    &        1.137     \\
\textbf{Atypicality\_of\_datasets} &       0.1464  &        0.046     &     3.210  &         0.001        &        0.057    &        0.236     \\
\textbf{NumDatasets\_log}          &      -0.1472  &        0.062     &    -2.372  &         0.018        &       -0.269    &       -0.026     \\
\textbf{Paper\_novelty}            &       0.1040  &        0.029     &     3.598  &         0.000        &        0.047    &        0.161     \\
\textbf{Data\_use\_frequency\_log} &       0.1613  &        0.029     &     5.651  &         0.000        &        0.105    &        0.217     \\
\textbf{NumAuthor\_log}            &       0.5047  &        0.062     &     8.181  &         0.000        &        0.384    &        0.626     \\
\textbf{AuthorExprience\_log}      &       0.0049  &        0.014     &     0.355  &         0.722        &       -0.022    &        0.032     \\
\end{tabular}
\caption{the Negative Binomial Regression Model Detail: the Effects of Atypicality of Data Combinations on 3-Year Citations for paper published between 1990 to 2000.}
\end{center}
\end{table*}

\begin{table*}
\begin{center}

  \renewcommand{\tablename}{Table S}
\tiny 

\begin{tabular}{lcccccc}
\tiny 
                                                  \textbf{Dep. Variable:}  3 year citation             & \textbf{coef} & \textbf{std err} & \textbf{z} & \textbf{P$> |$z$|$} & \textbf{[0.025} & \textbf{0.975]}  \\
\hline
\hline
\textbf{Intercept}                 &       1.3840  &        0.351     &     3.943  &         0.000        &        0.696    &        2.072     \\
\textbf{Atypicality\_of\_datasets} &       0.0919  &        0.081     &     1.138  &         0.255        &       -0.066    &        0.250     \\
\textbf{NumDatasets\_log}          &      -0.2347  &        0.120     &    -1.951  &         0.051        &       -0.470    &        0.001     \\
\textbf{Paper\_novelty}            &       0.1875  &        0.040     &     4.637  &         0.000        &        0.108    &        0.267     \\
\textbf{Data\_use\_frequency\_log} &      -0.0512  &        0.044     &    -1.165  &         0.244        &       -0.137    &        0.035     \\
\textbf{NumAuthor\_log}            &       0.3163  &        0.101     &     3.147  &         0.002        &        0.119    &        0.513     \\
\textbf{AuthorExprience\_log}      &       0.0318  &        0.023     &     1.358  &         0.175        &       -0.014    &        0.078     \\
\end{tabular}
\caption{the Negative Binomial Regression Model Detail: the Effects of Atypicality of Data Combinations on 3-Year Citations for paper published before 1990. }
    
\end{center}
\end{table*}

\FloatBarrier

\textbf{Impact of using atypical combinations of datasets on broader scientific impact - Wikipedia, Policy, News, and Twitter mentions.(Table 18 - 21)}

\FloatBarrier

\begin{table*}[ht]
  \renewcommand{\tablename}{Table S}
\begin{center}
\tiny 

\begin{tabular}{lcccccc}
\textbf{Dep. Variable:}  Wikipedia                                                              & \textbf{coef} & \textbf{std err} & \textbf{z} & \textbf{P$> |$z$|$} & \textbf{[0.025} & \textbf{0.975]}  \\
\hline
\hline
\textbf{Intercept}                      &      -4.7306  &        0.571     &    -8.285  &         0.000        &       -5.850    &       -3.611     \\
\textbf{Year\_bin[T.(2014, 2019]]} &       0.7344  &        0.150     &     4.882  &         0.000        &        0.440    &        1.029     \\
\textbf{Atypicality\_of\_datasets}      &       0.5655  &        0.106     &     5.340  &         0.000        &        0.358    &        0.773     \\
\textbf{Paper\_novelty}                 &      -0.1171  &        0.075     &    -1.554  &         0.120        &       -0.265    &        0.031     \\
\textbf{NumDatasets\_log}               &      -0.2935  &        0.132     &    -2.221  &         0.026        &       -0.552    &       -0.035     \\
\textbf{Data\_use\_frequency\_log}      &      -0.1375  &        0.059     &    -2.342  &         0.019        &       -0.253    &       -0.022     \\
\textbf{NumAuthor\_log}                 &       0.0341  &        0.126     &     0.272  &         0.786        &       -0.212    &        0.280     \\
\end{tabular}
\caption{
Results of the negative binomial regression table with number of Wikipedia mentions as the dependent variable and Atypicality of Data Combinations as independent variable. Since Altmetric began tracking data in 2011, we limit our broader impact analysis to papers published after 2010.} 
\end{center}
\end{table*}

\begin{table*}

\begin{center}
  \renewcommand{\tablename}{Table S}
\tiny 

\begin{tabular}{lcccccc}
 \textbf{Dep. Variable:} Policy                                                               & \textbf{coef} & \textbf{std err} & \textbf{z} & \textbf{P$> |$z$|$} & \textbf{[0.025} & \textbf{0.975]}  \\
\hline
\hline
\textbf{Intercept}                      &      -3.0098  &        0.398     &    -7.568  &         0.000        &       -3.789    &       -2.230     \\
\textbf{Year\_bin[T.(2014, 2019]]} &       0.0128  &        0.095     &     0.135  &         0.893        &       -0.174    &        0.199     \\
\textbf{Atypicality\_of\_datasets}      &       0.3691  &        0.072     &     5.121  &         0.000        &        0.228    &        0.510     \\
\textbf{Paper\_novelty}                 &       0.1301  &        0.060     &     2.167  &         0.030        &        0.012    &        0.248     \\
\textbf{NumDatasets\_log}               &      -0.2442  &        0.092     &    -2.662  &         0.008        &       -0.424    &       -0.064     \\
\textbf{Data\_use\_frequency\_log}      &      -0.2191  &        0.040     &    -5.483  &         0.000        &       -0.297    &       -0.141     \\
\textbf{NumAuthor\_log}                 &       0.5070  &        0.087     &     5.837  &         0.000        &        0.337    &        0.677     \\
\end{tabular}
\caption{Results of the negative binomial regression table with number of Policy document mentions as the dependent variable and Atypicality of Data Combinations as independent variable. Since Altmetric began tracking data in 2011, we limit our broader impact analysis to papers published after 2010.} 
\end{center}
\end{table*}

\begin{table*}

\begin{center}
  \renewcommand{\tablename}{Table S}
\tiny 

\begin{tabular}{lcccccc}
\textbf{Dep. Variable:} Twitter                                                               & \textbf{coef} & \textbf{std err} & \textbf{z} & \textbf{P$> |$z$|$} & \textbf{[0.025} & \textbf{0.975]}  \\
\hline
\hline

\textbf{Intercept}                      &      -0.0343  &        0.150     &    -0.229  &         0.819        &       -0.328    &        0.259     \\
\textbf{Year\_bin[T.(2014, 2019]]} &       1.5874  &        0.037     &    43.280  &         0.000        &        1.516    &        1.659     \\
\textbf{Atypicality\_of\_datasets}      &       0.5200  &        0.029     &    17.951  &         0.000        &        0.463    &        0.577     \\
\textbf{Paper\_novelty}                 &      -0.0434  &        0.023     &    -1.914  &         0.056        &       -0.088    &        0.001     \\
\textbf{NumDatasets\_log}               &      -0.1660  &        0.037     &    -4.457  &         0.000        &       -0.239    &       -0.093     \\
\textbf{Data\_use\_frequency\_log}      &      -0.2573  &        0.015     &   -16.653  &         0.000        &       -0.288    &       -0.227     \\
\textbf{NumAuthor\_log}                 &       0.3237  &        0.033     &     9.760  &         0.000        &        0.259    &        0.389     \\
\end{tabular}
\caption{Results of the Negative Binomial  regression table with number of Twitter mentions as the dependent variable and Atypicality of Data Combinations as independent variable. Since Altmetric began tracking data in 2011, we limit our broader impact analysis to papers published after 2010. } 
\end{center}
\end{table*}

\begin{table*}
\begin{center}
  \renewcommand{\tablename}{Table S}
\tiny 

\begin{tabular}{lcccccc}
 \textbf{Dep. Variable:}   News mention                                                            & \textbf{coef} & \textbf{std err} & \textbf{z} & \textbf{P$> |$z$|$} & \textbf{[0.025} & \textbf{0.975]}  \\

\hline
\hline
\textbf{Intercept}                      &      -3.0682  &        0.183     &   -16.764  &         0.000        &       -3.427    &       -2.709     \\
\textbf{Year\_bin[T.(2014, 2019]]} &       1.3452  &        0.044     &    30.686  &         0.000        &        1.259    &        1.431     \\
\textbf{Atypicality\_of\_datasets}      &       0.0858  &        0.034     &     2.551  &         0.011        &        0.020    &        0.152     \\
\textbf{NumDatasets\_log}               &       0.3658  &        0.042     &     8.661  &         0.000        &        0.283    &        0.449     \\
\textbf{Paper\_novelty}                 &      -0.0360  &        0.026     &    -1.389  &         0.165        &       -0.087    &        0.015     \\
\textbf{Data\_use\_frequency\_log}      &      -0.1338  &        0.018     &    -7.410  &         0.000        &       -0.169    &       -0.098     \\
\textbf{NumAuthor\_log}                 &       0.3682  &        0.039     &     9.522  &         0.000        &        0.292    &        0.444     \\
\end{tabular}
\caption{Results of the Negative Binomial regression table with number of News mentions as the dependent variable and Atypicality of Data Combinations as independent variable. Since Altmetric began tracking data in 2011, we limit our broader impact analysis to papers published after 2010. } 
\end{center}
\end{table*}

\FloatBarrier

\textbf{-Alternative Impact Quantification: We examined the impact of using atypical combinations of datasets on the likelihood of becoming top 5\% hit papers – publications that received citations within the top 5\% in our dataset.(Table 22)}

\FloatBarrier

\begin{table*}[ht]
\begin{center}
  \renewcommand{\tablename}{Table S}
\tiny 

\begin{tabular}{lcccccc}
\tiny 
\textbf{Dep. Variable:}  top 5\% hit paper (binary)  & \textbf{coef} & \textbf{std err} & \textbf{z} & \textbf{P$> |$z$|$} & \textbf{[0.025} & \textbf{0.975]}  \\
\hline
\hline
\textbf{Year\_bin[(1930, 1974]]} &     -25.6940  &     3.03e+04     &    -0.001  &         0.999        &    -5.95e+04    &     5.94e+04     \\
\textbf{Year\_bin[(1974, 1979]]} &     -26.7392  &     1.63e+04     &    -0.002  &         0.999        &     -3.2e+04    &      3.2e+04     \\
\textbf{Year\_bin[(1979, 1984]]} &      -8.0350  &        1.112     &    -7.226  &         0.000        &      -10.214    &       -5.856     \\
\textbf{Year\_bin[(1984, 1989]]} &      -8.9418  &        1.134     &    -7.888  &         0.000        &      -11.164    &       -6.720     \\
\textbf{Year\_bin[(1989, 1994]]} &      -7.6646  &        0.634     &   -12.092  &         0.000        &       -8.907    &       -6.422     \\
\textbf{Year\_bin[(1994, 1999]]} &      -7.1479  &        0.513     &   -13.942  &         0.000        &       -8.153    &       -6.143     \\
\textbf{Year\_bin[(1999, 2004]]} &      -6.7771  &        0.502     &   -13.489  &         0.000        &       -7.762    &       -5.792     \\
\textbf{Year\_bin[(2004, 2009]]} &      -6.8618  &        0.508     &   -13.506  &         0.000        &       -7.858    &       -5.866     \\
\textbf{Year\_bin[(2009, 2014]]} &      -6.9779  &        0.501     &   -13.938  &         0.000        &       -7.959    &       -5.997     \\
\textbf{Year\_bin[(2014, 2019]]} &      -7.3148  &        0.505     &   -14.489  &         0.000        &       -8.304    &       -6.325     \\
\textbf{Atypicality\_of\_datasets}    &       0.4164  &        0.079     &     5.253  &         0.000        &        0.261    &        0.572     \\
\textbf{NumDatasets\_log}             &      -0.2890  &        0.110     &    -2.617  &         0.009        &       -0.505    &       -0.073     \\
\textbf{Paper\_novelty}               &       0.2851  &        0.073     &     3.904  &         0.000        &        0.142    &        0.428     \\
\textbf{Data\_use\_frequency\_log}    &       0.1826  &        0.049     &     3.745  &         0.000        &        0.087    &        0.278     \\
\textbf{NumAuthor\_log}               &       0.8264  &        0.098     &     8.448  &         0.000        &        0.635    &        1.018     \\
\textbf{AuthorExprience\_log}         &      -0.0658  &        0.031     &    -2.132  &         0.033        &       -0.126    &       -0.005     \\
\textbf{ImpactFactor\_log}            &       1.1438  &        0.095     &    11.991  &         0.000        &        0.957    &        1.331     \\
\textbf{Biology}                      &       1.6209  &        1.460     &     1.110  &         0.267        &       -1.242    &        4.483     \\
\textbf{Business}                     &      -0.9146  &        1.140     &    -0.802  &         0.422        &       -3.149    &        1.320     \\
\textbf{Chemistry}                    &       0.5847  &        1.446     &     0.404  &         0.686        &       -2.250    &        3.419     \\
\textbf{Computer\_science}            &      -0.8423  &        1.738     &    -0.485  &         0.628        &       -4.249    &        2.565     \\
\textbf{Economics}                    &       1.2835  &        0.560     &     2.292  &         0.022        &        0.186    &        2.381     \\
\textbf{Engineering}                  &      -1.5627  &        3.001     &    -0.521  &         0.603        &       -7.445    &        4.320     \\
\textbf{Environmental\_science}       &       0.0144  &        2.459     &     0.006  &         0.995        &       -4.805    &        4.834     \\
\textbf{Geography}                    &      -0.5714  &        0.862     &    -0.663  &         0.507        &       -2.261    &        1.118     \\
\textbf{History}                      &       2.0976  &        2.666     &     0.787  &         0.431        &       -3.127    &        7.323     \\
\textbf{Materials\_science}           &     -60.3677  &     3.69e+05     &    -0.000  &         1.000        &    -7.23e+05    &     7.23e+05     \\
\textbf{Mathematics}                  &      -0.0353  &        1.924     &    -0.018  &         0.985        &       -3.807    &        3.736     \\
\textbf{Medicine}                     &       1.4774  &        0.344     &     4.293  &         0.000        &        0.803    &        2.152     \\
\textbf{Philosophy}                   &     -52.8283  &     1.79e+05     &    -0.000  &         1.000        &    -3.51e+05    &     3.51e+05     \\
\textbf{Political\_science}           &      -1.8446  &        0.771     &    -2.392  &         0.017        &       -3.356    &       -0.333     \\
\textbf{Psychology}                   &       0.1554  &        0.394     &     0.395  &         0.693        &       -0.616    &        0.927     \\
\textbf{Sociology}                    &       1.9738  &        0.685     &     2.882  &         0.004        &        0.631    &        3.316     \\
\hline

\textbf{  No. Observations:  } &     8836      & \textbf{  Log-Likelihood:    } &    -1456.2    \\
\textbf{  Df Model:          } &       32     & \textbf{  Df Residuals:      } &     8803    \\
 \textbf{  Pearson chi2:      } &  1.30e+04  & \textbf{  Deviance:          } &    2912.4 \\
\end{tabular}
\caption{Logistic regression model: investigating the impact of atypicality of data combinations on achieving top 5 percent hit paper status. This study serves as a robustness test. The hit paper variable is binary, with 1 indicating that the publication received citations in the top 5 percent among all the papers in our dataset, and 0 otherwise. }
\end{center}
\end{table*}

\FloatBarrier
\newpage
\textbf{(3) The effect of atypical dataset topic combinations on
scientific impact: -Impact of using atypical dataset topic combinations and atypical combinations of datasets on citation over 3, 5, 10 year. (Table 23-25)}

\FloatBarrier

\begin{table*}[ht]
\begin{center}
\tiny 
  \renewcommand{\tablename}{Table S}
\begin{tabular}{lcccccc}
\textbf{Dep. Variable:} 3 year citation                                                                    & \textbf{coef} & \textbf{std err} & \textbf{z} & \textbf{P$> |$z$|$} & \textbf{[0.025} & \textbf{0.975]}  \\
 \hline
 \hline
\textbf{Intercept}                      &      -0.0844  &        0.215     &    -0.393  &         0.695        &       -0.506    &        0.337     \\
\textbf{Year\_bin[T.(1974, 1979]]} &       0.0446  &        0.232     &     0.192  &         0.848        &       -0.410    &        0.500     \\
\textbf{Year\_bin[T.(1979, 1984]]} &       0.2642  &        0.216     &     1.221  &         0.222        &       -0.160    &        0.688     \\
\textbf{Year\_bin[T.(1984, 1989]]} &       0.2446  &        0.214     &     1.145  &         0.252        &       -0.174    &        0.663     \\
\textbf{Year\_bin[T.(1989, 1994]]} &       0.4423  &        0.212     &     2.083  &         0.037        &        0.026    &        0.859     \\
\textbf{Year\_bin[T.(1994, 1999]]} &       0.8235  &        0.211     &     3.903  &         0.000        &        0.410    &        1.237     \\
\textbf{Year\_bin[T.(1999, 2004]]} &       0.9114  &        0.211     &     4.324  &         0.000        &        0.498    &        1.325     \\
\textbf{Year\_bin[T.(2004, 2009]]} &       1.0503  &        0.211     &     4.987  &         0.000        &        0.638    &        1.463     \\
\textbf{Year\_bin[T.(2009, 2014]]} &       0.9407  &        0.211     &     4.468  &         0.000        &        0.528    &        1.353     \\
\textbf{Year\_bin[T.(2014, 2019]]} &       0.7320  &        0.211     &     3.476  &         0.001        &        0.319    &        1.145     \\
\textbf{Atypicality\_of\_datasets}      &       0.1614  &        0.010     &    16.088  &         0.000        &        0.142    &        0.181     \\
\textbf{Topic\_atypicality}             &       0.0389  &        0.007     &     5.465  &         0.000        &        0.025    &        0.053     \\
\textbf{NumDatasets\_log}               &      -0.1009  &        0.013     &    -7.504  &         0.000        &       -0.127    &       -0.075     \\
\textbf{Paper\_novelty}                 &       0.1437  &        0.007     &    19.357  &         0.000        &        0.129    &        0.158     \\
\textbf{Data\_use\_frequency\_log}      &       0.0955  &        0.006     &    16.934  &         0.000        &        0.084    &        0.107     \\
\textbf{NumAuthor\_log}                 &       0.3250  &        0.012     &    27.874  &         0.000        &        0.302    &        0.348     \\
\textbf{AuthorExprience\_log}           &       0.0053  &        0.003     &     1.550  &         0.121        &       -0.001    &        0.012     \\
\textbf{ImpactFactor\_log}              &       0.4620  &        0.010     &    45.510  &         0.000        &        0.442    &        0.482     \\
\textbf{Biology}                        &      -0.0446  &        0.256     &    -0.174  &         0.862        &       -0.546    &        0.457     \\
\textbf{Business}                       &       0.2082  &        0.077     &     2.711  &         0.007        &        0.058    &        0.359     \\
\textbf{Chemistry}                      &       0.6722  &        0.230     &     2.925  &         0.003        &        0.222    &        1.122     \\
\textbf{Computer\_science}              &       0.1860  &        0.124     &     1.499  &         0.134        &       -0.057    &        0.429     \\
\textbf{Economics}                      &       0.5935  &        0.056     &    10.617  &         0.000        &        0.484    &        0.703     \\
\textbf{Engineering}                    &      -1.0141  &        0.181     &    -5.591  &         0.000        &       -1.370    &       -0.659     \\
\textbf{Environmental\_science}         &       0.2277  &        0.251     &     0.907  &         0.364        &       -0.264    &        0.720     \\
\textbf{Geography}                      &       0.3010  &        0.074     &     4.053  &         0.000        &        0.155    &        0.447     \\
\textbf{History}                        &       0.0139  &        0.242     &     0.057  &         0.954        &       -0.460    &        0.488     \\
\textbf{Materials\_science}             &      -0.2205  &        1.583     &    -0.139  &         0.889        &       -3.324    &        2.883     \\
\textbf{Mathematics}                    &      -0.8689  &        0.194     &    -4.476  &         0.000        &       -1.249    &       -0.488     \\
\textbf{Medicine}                       &       0.8285  &        0.037     &    22.197  &         0.000        &        0.755    &        0.902     \\
\textbf{Philosophy}                     &      -0.4879  &        1.432     &    -0.341  &         0.733        &       -3.294    &        2.319     \\
\textbf{Political\_science}             &      -0.0462  &        0.053     &    -0.866  &         0.386        &       -0.151    &        0.058     \\
\textbf{Psychology}                     &       0.1281  &        0.042     &     3.079  &         0.002        &        0.047    &        0.210     \\
\textbf{Sociology}                      &       0.4639  &        0.061     &     7.572  &         0.000        &        0.344    &        0.584     \\
\hline
  \textbf{  No. Observations:  } &     8812         & \textbf{  Log-Likelihood:    } &    -36932. & \textbf{  Dispersion:    } &  0.25 \\
\textbf{  Df Residuals:      } &     8778     & \textbf{  Df Model:          } &       33    \\
\textbf{  Pearson chi2:      } &  8.10e+04 & \textbf{  Deviance:          } &    28111.  \\

\end{tabular}
 \caption{
 Results of the Negative Binomial regression table with 3-Year Citations as the dependent variable and Atypicality of Data Combinations and Topic Atypicality as independent variables. } 
\end{center}
\end{table*}

\begin{table*}
\begin{center}
  \renewcommand{\tablename}{Table S}
\tiny 

\begin{tabular}{lcccccc}
\tiny 
                                                  \textbf{Dep. Variable:}  5 year citation             & \textbf{coef} & \textbf{std err} & \textbf{z} & \textbf{P$> |$z$|$} & \textbf{[0.025} & \textbf{0.975]}  \\
\hline
\hline
\textbf{Intercept}                      &       0.8266  &        0.182     &     4.544  &         0.000        &        0.470    &        1.183     \\
\textbf{Year\_bin[T.(1974, 1979]]} &       0.1574  &        0.195     &     0.806  &         0.420        &       -0.225    &        0.540     \\
\textbf{Year\_bin[T.(1979, 1984]]} &       0.1937  &        0.182     &     1.062  &         0.288        &       -0.164    &        0.551     \\
\textbf{Year\_bin[T.(1984, 1989]]} &       0.1832  &        0.180     &     1.019  &         0.308        &       -0.169    &        0.536     \\
\textbf{Year\_bin[T.(1989, 1994]]} &       0.4079  &        0.179     &     2.282  &         0.022        &        0.058    &        0.758     \\
\textbf{Year\_bin[T.(1994, 1999]]} &       0.8974  &        0.177     &     5.060  &         0.000        &        0.550    &        1.245     \\
\textbf{Year\_bin[T.(1999, 2004]]} &       0.9250  &        0.177     &     5.221  &         0.000        &        0.578    &        1.272     \\
\textbf{Year\_bin[T.(2004, 2009]]} &       1.0257  &        0.177     &     5.794  &         0.000        &        0.679    &        1.373     \\
\textbf{Year\_bin[T.(2009, 2014]]} &       0.8454  &        0.177     &     4.778  &         0.000        &        0.499    &        1.192     \\
\textbf{Year\_bin[T.(2014, 2019]]} &       0.6615  &        0.177     &     3.732  &         0.000        &        0.314    &        1.009     \\
\textbf{Atypicality\_of\_datasets}      &       0.1666  &        0.010     &    16.743  &         0.000        &        0.147    &        0.186     \\
\textbf{Topic\_atypicality}             &       0.0354  &        0.007     &     4.895  &         0.000        &        0.021    &        0.050     \\
\textbf{Paper\_novelty}                 &       0.1685  &        0.007     &    23.321  &         0.000        &        0.154    &        0.183     \\
\textbf{NumDatasets\_log}               &      -0.1380  &        0.013     &   -10.264  &         0.000        &       -0.164    &       -0.112     \\
\textbf{Data\_use\_frequency\_log}      &       0.0831  &        0.006     &    14.687  &         0.000        &        0.072    &        0.094     \\
\textbf{NumAuthor\_log}                 &       0.3302  &        0.012     &    28.398  &         0.000        &        0.307    &        0.353     \\
\textbf{AuthorExprience\_log}           &      -0.0065  &        0.003     &    -1.940  &         0.052        &       -0.013    &     6.58e-05     \\
\textbf{ImpactFactor\_log}              &       0.4265  &        0.010     &    43.454  &         0.000        &        0.407    &        0.446     \\
\textbf{Biology}                        &       0.2384  &        0.247     &     0.964  &         0.335        &       -0.246    &        0.723     \\
\textbf{Business}                       &       0.0621  &        0.077     &     0.810  &         0.418        &       -0.088    &        0.212     \\
\textbf{Chemistry}                      &       0.6941  &        0.226     &     3.066  &         0.002        &        0.250    &        1.138     \\
\textbf{Computer\_science}              &       0.3353  &        0.122     &     2.753  &         0.006        &        0.097    &        0.574     \\
\textbf{Economics}                      &       0.4822  &        0.054     &     8.872  &         0.000        &        0.376    &        0.589     \\
\textbf{Engineering}                    &      -0.9814  &        0.169     &    -5.818  &         0.000        &       -1.312    &       -0.651     \\
\textbf{Environmental\_science}         &       0.2388  &        0.242     &     0.987  &         0.324        &       -0.235    &        0.713     \\
\textbf{Geography}                      &       0.2328  &        0.073     &     3.208  &         0.001        &        0.091    &        0.375     \\
\textbf{History}                        &       0.0138  &        0.234     &     0.059  &         0.953        &       -0.444    &        0.472     \\
\textbf{Materials\_science}             &       1.1074  &        1.491     &     0.743  &         0.458        &       -1.815    &        4.030     \\
\textbf{Mathematics}                    &      -0.7635  &        0.181     &    -4.218  &         0.000        &       -1.118    &       -0.409     \\
\textbf{Medicine}                       &       0.8105  &        0.037     &    21.899  &         0.000        &        0.738    &        0.883     \\
\textbf{Philosophy}                     &       2.2116  &        0.992     &     2.229  &         0.026        &        0.267    &        4.156     \\
\textbf{Political\_science}             &      -0.1193  &        0.052     &    -2.306  &         0.021        &       -0.221    &       -0.018     \\
\textbf{Psychology}                     &       0.1717  &        0.041     &     4.152  &         0.000        &        0.091    &        0.253     \\
\textbf{Sociology}                      &       0.5920  &        0.059     &    10.026  &         0.000        &        0.476    &        0.708     \\
\hline
\textbf{  No. Observations:  } &     7956  & \textbf{  Log-Likelihood:    } &  -38761.  & \textbf{  Dispersion:    } &  0.25 \\
\textbf{  Df Model:          } &       33     & \textbf{  Df Residuals:      } &      7922    \\
 \textbf{  Pearson chi2:      } &  7.35e+04    & \textbf{  Deviance:          } &   27893. \\

\end{tabular}
\end{center}
\caption{the Negative Binomial Regression Model Detail: the Effects of Atypicality of Data Combinations and Topic Atypicality on 5-Year Citations. {\it Note:} This study serves as a robustness test. To capture 5-year citations, we track publications in our dataset up to 2018 for this analysis. }
\end{table*}

\begin{table*}
\begin{center}
  \renewcommand{\tablename}{Table S}
\tiny 

\begin{tabular}{lcccccc}
\tiny 
                                                  \textbf{Dep. Variable:}  10 year citation             & \textbf{coef} & \textbf{std err} & \textbf{z} & \textbf{P$> |$z$|$} & \textbf{[0.025} & \textbf{0.975]}  \\
\hline
\hline
\textbf{Intercept}                      &       1.8362  &        0.161     &    11.404  &         0.000        &        1.521    &        2.152     \\
\textbf{Year\_bin[T.(1974, 1979]]} &       0.0455  &        0.171     &     0.266  &         0.790        &       -0.290    &        0.381     \\
\textbf{Year\_bin[T.(1979, 1984]]} &       0.0707  &        0.159     &     0.444  &         0.657        &       -0.241    &        0.383     \\
\textbf{Year\_bin[T.(1984, 1989]]} &       0.1174  &        0.157     &     0.750  &         0.453        &       -0.190    &        0.424     \\
\textbf{Year\_bin[T.(1989, 1994]]} &       0.4794  &        0.156     &     3.080  &         0.002        &        0.174    &        0.785     \\
\textbf{Year\_bin[T.(1994, 1999]]} &       0.9962  &        0.154     &     6.450  &         0.000        &        0.693    &        1.299     \\
\textbf{Year\_bin[T.(1999, 2004]]} &       1.0153  &        0.154     &     6.580  &         0.000        &        0.713    &        1.318     \\
\textbf{Year\_bin[T.(2004, 2009]]} &       0.9699  &        0.154     &     6.289  &         0.000        &        0.668    &        1.272     \\
\textbf{Year\_bin[T.(2009, 2014]]} &       0.7941  &        0.154     &     5.144  &         0.000        &        0.492    &        1.097     \\
\textbf{Atypicality\_of\_datasets}      &       0.1685  &        0.011     &    15.275  &         0.000        &        0.147    &        0.190     \\
\textbf{Topic\_atypicality}             &       0.0426  &        0.008     &     5.081  &         0.000        &        0.026    &        0.059     \\
\textbf{Paper\_novelty}                 &       0.1820  &        0.008     &    23.661  &         0.000        &        0.167    &        0.197     \\
\textbf{NumDatasets\_log}               &      -0.1698  &        0.015     &   -11.111  &         0.000        &       -0.200    &       -0.140     \\
\textbf{Data\_use\_frequency\_log}      &       0.0639  &        0.006     &    10.042  &         0.000        &        0.051    &        0.076     \\
\textbf{NumAuthor\_log}                 &       0.3574  &        0.013     &    26.925  &         0.000        &        0.331    &        0.383     \\
\textbf{AuthorExprience\_log}           &      -0.0050  &        0.004     &    -1.372  &         0.170        &       -0.012    &        0.002     \\
\textbf{ImpactFactor\_log}              &       0.3727  &        0.010     &    35.740  &         0.000        &        0.352    &        0.393     \\
\textbf{Biology}                        &       0.4787  &        0.282     &     1.696  &         0.090        &       -0.075    &        1.032     \\
\textbf{Business}                       &      -0.3313  &        0.083     &    -3.972  &         0.000        &       -0.495    &       -0.168     \\
\textbf{Chemistry}                      &       0.8378  &        0.234     &     3.583  &         0.000        &        0.380    &        1.296     \\
\textbf{Computer\_science}              &      -0.1293  &        0.147     &    -0.880  &         0.379        &       -0.417    &        0.159     \\
\textbf{Economics}                      &       0.5471  &        0.058     &     9.509  &         0.000        &        0.434    &        0.660     \\
\textbf{Engineering}                    &      -1.3495  &        0.172     &    -7.833  &         0.000        &       -1.687    &       -1.012     \\
\textbf{Environmental\_science}         &       0.0239  &        0.277     &     0.086  &         0.931        &       -0.520    &        0.567     \\
\textbf{Geography}                      &       0.1294  &        0.080     &     1.613  &         0.107        &       -0.028    &        0.287     \\
\textbf{History}                        &      -0.7160  &        0.251     &    -2.858  &         0.004        &       -1.207    &       -0.225     \\
\textbf{Mathematics}                    &      -0.5565  &        0.187     &    -2.982  &         0.003        &       -0.922    &       -0.191     \\
\textbf{Medicine}                       &       0.7184  &        0.041     &    17.418  &         0.000        &        0.638    &        0.799     \\
\textbf{Philosophy}                     &       0.6067  &        0.937     &     0.647  &         0.517        &       -1.231    &        2.444     \\
\textbf{Political\_science}             &      -0.3076  &        0.055     &    -5.582  &         0.000        &       -0.416    &       -0.200     \\
\textbf{Psychology}                     &       0.2479  &        0.047     &     5.288  &         0.000        &        0.156    &        0.340     \\
\textbf{Sociology}                      &       0.8343  &        0.063     &    13.150  &         0.000        &        0.710    &        0.959     \\
\hline
\textbf{  No. Observations:  } &     5801  & \textbf{  Log-Likelihood:    } &  -34140.   & \textbf{  Dispersion:    } &  0.25 \\
\textbf{  Df Model:          } &      31     & \textbf{  Df Residuals:      } &      5769    \\
 \textbf{  Pearson chi2:      } &   5.94e+04    & \textbf{  Deviance:          } &   24360.  \\

\end{tabular}
\caption{the Negative Binomial Regression Model Detail: the Effects of Atypicality of Data Combinations and Topic Atypicality on 10-Year Citations. {\it Note:} This study serves as a robustness test. To capture 10-year citations, we track publications in our dataset up to 2013 for this analysis. }
\end{center}
\end{table*}

\FloatBarrier

\textbf{Tables 26-29 present the effects of using atypical combinations of datasets on citations, based on a three-year analysis of publications released in four distinct time periods: before 1990, 1990-2000, 2000-2010, and 2010-2020.}

\FloatBarrier

\begin{table*}[ht]
\begin{center}
  \renewcommand{\tablename}{Table S}
\tiny 

\begin{tabular}{lcccccc}
\tiny 
                                                  \textbf{Dep. Variable:}  3 year citation             & \textbf{coef} & \textbf{std err} & \textbf{z} & \textbf{P$> |$z$|$} & \textbf{[0.025} & \textbf{0.975]}  \\
\hline
\hline
\textbf{Intercept}                 &       0.6385  &        0.132     &     4.829  &         0.000        &        0.379    &        0.898     \\
\textbf{Atypicality\_of\_datasets} &       0.1801  &        0.027     &     6.786  &         0.000        &        0.128    &        0.232     \\
\textbf{Topic\_atypicality}        &       0.0583  &        0.018     &     3.318  &         0.001        &        0.024    &        0.093     \\
\textbf{NumDatasets\_log}          &      -0.0751  &        0.034     &    -2.177  &         0.029        &       -0.143    &       -0.007     \\
\textbf{Paper\_novelty}            &       0.1423  &        0.020     &     6.943  &         0.000        &        0.102    &        0.182     \\
\textbf{Data\_use\_frequency\_log} &       0.1067  &        0.014     &     7.522  &         0.000        &        0.079    &        0.134     \\
\textbf{NumAuthor\_log}            &       0.3016  &        0.029     &    10.432  &         0.000        &        0.245    &        0.358     \\
\textbf{AuthorExprience\_log}      &       0.0037  &        0.009     &     0.415  &         0.678        &       -0.014    &        0.021     \\
\textbf{ImpactFactor\_log}         &       0.5080  &        0.027     &    19.118  &         0.000        &        0.456    &        0.560     \\
\textbf{Biology}                   &      -1.0409  &        0.582     &    -1.789  &         0.074        &       -2.181    &        0.099     \\
\textbf{Business}                  &       0.2570  &        0.199     &     1.290  &         0.197        &       -0.134    &        0.647     \\
\textbf{Chemistry}                 &       0.7640  &        0.528     &     1.448  &         0.148        &       -0.270    &        1.798     \\
\textbf{Computer\_science}         &       0.2594  &        0.293     &     0.887  &         0.375        &       -0.314    &        0.833     \\
\textbf{Economics}                 &       0.4139  &        0.168     &     2.458  &         0.014        &        0.084    &        0.744     \\
\textbf{Engineering}               &      -0.5083  &        0.532     &    -0.955  &         0.340        &       -1.551    &        0.535     \\
\textbf{Environmental\_science}    &       0.7519  &        0.610     &     1.233  &         0.218        &       -0.443    &        1.947     \\
\textbf{Geography}                 &      -0.1736  &        0.198     &    -0.875  &         0.382        &       -0.562    &        0.215     \\
\textbf{History}                   &       0.3999  &        0.638     &     0.627  &         0.531        &       -0.851    &        1.651     \\
\textbf{Materials\_science}        &      -0.6143  &        2.968     &    -0.207  &         0.836        &       -6.431    &        5.203     \\
\textbf{Mathematics}               &      -0.5090  &        0.491     &    -1.037  &         0.300        &       -1.471    &        0.453     \\
\textbf{Medicine}                  &       0.7960  &        0.097     &     8.206  &         0.000        &        0.606    &        0.986     \\
\textbf{Political\_science}        &       0.2125  &        0.150     &     1.420  &         0.156        &       -0.081    &        0.506     \\
\textbf{Psychology}                &       0.1448  &        0.105     &     1.381  &         0.167        &       -0.061    &        0.350     \\
\textbf{Sociology}                 &       0.1779  &        0.169     &     1.050  &         0.294        &       -0.154    &        0.510     \\
\hline
\textbf{  No. Observations:  } &     4430  & \textbf{  Log-Likelihood:    } & -16673. & \textbf{  Dispersion:    } &  1  \\
\textbf{  Df Model:          } &      23     & \textbf{  Df Residuals:      } &      4406    \\
 \textbf{  Pearson chi2:      } & 1.31e+04    & \textbf{  Deviance:          } &    4591.4\\

\end{tabular}
\caption{the Negative Binomial Regression Model Detail: the Effects of Atypicality of Data Combinations and Topic Atypicality on 3-Year Citations for paper published between 2010 and 2020.  }
\end{center}
\end{table*}

\begin{table*}
\begin{center}
  \renewcommand{\tablename}{Table S}
\tiny 

\begin{tabular}{lcccccc}
\tiny 
                                                  \textbf{Dep. Variable:}  3 year citation             & \textbf{coef} & \textbf{std err} & \textbf{z} & \textbf{P$> |$z$|$} & \textbf{[0.025} & \textbf{0.975]}  \\
\hline
\hline
\textbf{Intercept}                 &       1.0901  &        0.149     &     7.306  &         0.000        &        0.798    &        1.383     \\
\textbf{Atypicality\_of\_datasets} &       0.1884  &        0.029     &     6.411  &         0.000        &        0.131    &        0.246     \\
\textbf{Topic\_atypicality}        &       0.0200  &        0.023     &     0.862  &         0.389        &       -0.025    &        0.065     \\
\textbf{NumDatasets\_log}          &      -0.1556  &        0.041     &    -3.785  &         0.000        &       -0.236    &       -0.075     \\
\textbf{Paper\_novelty}            &       0.1681  &        0.022     &     7.780  &         0.000        &        0.126    &        0.210     \\
\textbf{Data\_use\_frequency\_log} &       0.0919  &        0.017     &     5.476  &         0.000        &        0.059    &        0.125     \\
\textbf{NumAuthor\_log}            &       0.2888  &        0.035     &     8.350  &         0.000        &        0.221    &        0.357     \\
\textbf{AuthorExprience\_log}      &       0.0079  &        0.010     &     0.763  &         0.446        &       -0.012    &        0.028     \\
\textbf{ImpactFactor\_log}         &       0.4141  &        0.028     &    14.938  &         0.000        &        0.360    &        0.468     \\
\textbf{Biology}                   &       0.9429  &        0.733     &     1.286  &         0.199        &       -0.495    &        2.380     \\
\textbf{Business}                  &       0.2241  &        0.215     &     1.043  &         0.297        &       -0.197    &        0.645     \\
\textbf{Chemistry}                 &       0.3879  &        0.688     &     0.564  &         0.573        &       -0.960    &        1.736     \\
\textbf{Computer\_science}         &       0.1309  &        0.423     &     0.309  &         0.757        &       -0.699    &        0.961     \\
\textbf{Economics}                 &       0.6101  &        0.165     &     3.693  &         0.000        &        0.286    &        0.934     \\
\textbf{Engineering}               &      -1.7452  &        0.470     &    -3.714  &         0.000        &       -2.666    &       -0.824     \\
\textbf{Environmental\_science}    &      -0.0074  &        0.674     &    -0.011  &         0.991        &       -1.328    &        1.313     \\
\textbf{Geography}                 &       0.3082  &        0.213     &     1.447  &         0.148        &       -0.109    &        0.725     \\
\textbf{History}                   &      -0.6222  &        0.683     &    -0.911  &         0.362        &       -1.960    &        0.716     \\
\textbf{Mathematics}               &      -0.8719  &        0.546     &    -1.597  &         0.110        &       -1.942    &        0.198     \\
\textbf{Medicine}                  &       1.0145  &        0.112     &     9.081  &         0.000        &        0.796    &        1.233     \\
\textbf{Philosophy}                &       1.9073  &        3.307     &     0.577  &         0.564        &       -4.574    &        8.388     \\
\textbf{Political\_science}        &      -0.1005  &        0.158     &    -0.637  &         0.524        &       -0.409    &        0.208     \\
\textbf{Psychology}                &       0.0988  &        0.127     &     0.777  &         0.437        &       -0.150    &        0.348     \\
\textbf{Sociology}                 &       0.6453  &        0.177     &     3.643  &         0.000        &        0.298    &        0.993     \\
\hline
\textbf{  No. Observations:  } &     3230  & \textbf{  Log-Likelihood:    } &  -12697. & \textbf{  Dispersion:    } &  1   \\
\textbf{  Df Model:          } &      23     & \textbf{  Df Residuals:      } &      3206   \\
 \textbf{  Pearson chi2:      } &  9.74e+03
    & \textbf{  Deviance:          } &     3356.3 \\

\end{tabular}
\end{center}
\caption{the Negative Binomial Regression Model Detail: the Effects of Atypicality of Data Combinations and Topic Atypicality on 3-Year Citations for paper published between 2000 and 2010.  }
\end{table*}

\begin{table*}
\begin{center}
  \renewcommand{\tablename}{Table S}
\tiny 
\begin{tabular}{lcccccc}
\tiny 
                                                  \textbf{Dep. Variable:}  3 year citation             & \textbf{coef} & \textbf{std err} & \textbf{z} & \textbf{P$> |$z$|$} & \textbf{[0.025} & \textbf{0.975]}  \\
\hline
\hline
\textbf{Intercept}                 &       0.5850  &        0.223     &     2.629  &         0.009        &        0.149    &        1.021     \\
\textbf{Atypicality\_of\_datasets} &       0.1170  &        0.048     &     2.462  &         0.014        &        0.024    &        0.210     \\
\textbf{Topic\_atypicality}        &       0.0633  &        0.035     &     1.832  &         0.067        &       -0.004    &        0.131     \\
\textbf{NumDatasets\_log}          &      -0.1335  &        0.062     &    -2.141  &         0.032        &       -0.256    &       -0.011     \\
\textbf{Paper\_novelty}            &       0.1013  &        0.029     &     3.505  &         0.000        &        0.045    &        0.158     \\
\textbf{Data\_use\_frequency\_log} &       0.1894  &        0.032     &     5.948  &         0.000        &        0.127    &        0.252     \\
\textbf{NumAuthor\_log}            &       0.5207  &        0.062     &     8.416  &         0.000        &        0.399    &        0.642     \\
\textbf{AuthorExprience\_log}      &       0.0027  &        0.014     &     0.193  &         0.847        &       -0.025    &        0.030     \\
\textbf{ImpactFactor\_log}         &       0.3363  &        0.045     &     7.543  &         0.000        &        0.249    &        0.424     \\
\textbf{Biology}                   &      -3.5703  &        3.012     &    -1.185  &         0.236        &       -9.474    &        2.333     \\
\textbf{Business}                  &       0.4638  &        0.335     &     1.383  &         0.167        &       -0.194    &        1.121     \\
\textbf{Computer\_science}         &      -0.8772  &        0.636     &    -1.380  &         0.168        &       -2.123    &        0.369     \\
\textbf{Economics}                 &       0.7439  &        0.206     &     3.619  &         0.000        &        0.341    &        1.147     \\
\textbf{Engineering}               &       0.0162  &        0.691     &     0.023  &         0.981        &       -1.338    &        1.370     \\
\textbf{Environmental\_science}    &      -5.2230  &        3.696     &    -1.413  &         0.158        &      -12.468    &        2.022     \\
\textbf{Geography}                 &       0.7504  &        0.338     &     2.219  &         0.026        &        0.088    &        1.413     \\
\textbf{History}                   &      -0.9929  &        0.977     &    -1.017  &         0.309        &       -2.907    &        0.921     \\
\textbf{Mathematics}               &      -2.3359  &        0.770     &    -3.032  &         0.002        &       -3.846    &       -0.826     \\
\textbf{Medicine}                  &       0.6046  &        0.167     &     3.617  &         0.000        &        0.277    &        0.932     \\
\textbf{Philosophy}                &      -0.1964  &        2.424     &    -0.081  &         0.935        &       -4.947    &        4.554     \\
\textbf{Political\_science}        &      -0.1671  &        0.214     &    -0.782  &         0.434        &       -0.586    &        0.252     \\
\textbf{Psychology}                &      -0.2268  &        0.195     &    -1.160  &         0.246        &       -0.610    &        0.156     \\
\textbf{Sociology}                 &       0.6786  &        0.242     &     2.808  &         0.005        &        0.205    &        1.152     \\
\hline
\textbf{  No. Observations:  } &     1306  & \textbf{  Log-Likelihood:    } &  -4458.7 & \textbf{  Dispersion:    } &  1  \\
\textbf{  Df Model:          } &      22     & \textbf{  Df Residuals:      } &      1283   \\
 \textbf{  Pearson chi2:      } &   2.93e+03   & \textbf{  Deviance:          } &     1583.8 \\

\end{tabular}
\end{center}
\caption{the Negative Binomial Regression Model Detail: the Effects of Atypicality of Data Combinations and Topic Atypicality on 3-Year Citations for paper published between 1990 and 2000.  }
\end{table*}

\begin{table*}
\begin{center}
  \renewcommand{\tablename}{Table S}
\tiny 
\begin{tabular}{lcccccc}
\tiny 
                                                  \textbf{Dep. Variable:}  3 year citation             & \textbf{coef} & \textbf{std err} & \textbf{z} & \textbf{P$> |$z$|$} & \textbf{[0.025} & \textbf{0.975]}  \\
\hline
\hline
\textbf{Intercept}                 &       1.4028  &        0.363     &     3.869  &         0.000        &        0.692    &        2.114     \\
\textbf{Atypicality\_of\_datasets} &       0.0951  &        0.083     &     1.152  &         0.249        &       -0.067    &        0.257     \\
\textbf{Topic\_atypicality}        &      -0.0157  &        0.058     &    -0.271  &         0.787        &       -0.129    &        0.098     \\
\textbf{NumDatasets\_log}          &      -0.2319  &        0.120     &    -1.928  &         0.054        &       -0.468    &        0.004     \\
\textbf{Paper\_novelty}            &       0.1872  &        0.040     &     4.626  &         0.000        &        0.108    &        0.267     \\
\textbf{Data\_use\_frequency\_log} &      -0.0588  &        0.052     &    -1.130  &         0.259        &       -0.161    &        0.043     \\
\textbf{NumAuthor\_log}            &       0.3165  &        0.101     &     3.142  &         0.002        &        0.119    &        0.514     \\
\textbf{AuthorExprience\_log}      &       0.0328  &        0.024     &     1.393  &         0.163        &       -0.013    &        0.079     \\
\textbf{ImpactFactor\_log}         &      -0.1144  &        0.076     &    -1.497  &         0.134        &       -0.264    &        0.035     \\
\textbf{Business}                  &      -0.2508  &        0.583     &    -0.430  &         0.667        &       -1.393    &        0.892     \\
\textbf{Chemistry}                 &       1.4315  &        2.324     &     0.616  &         0.538        &       -3.123    &        5.987     \\
\textbf{Computer\_science}         &      -0.1612  &        0.759     &    -0.212  &         0.832        &       -1.649    &        1.327     \\
\textbf{Economics}                 &       0.8655  &        0.328     &     2.640  &         0.008        &        0.223    &        1.508     \\
\textbf{Engineering}               &      -0.9989  &        1.051     &    -0.950  &         0.342        &       -3.059    &        1.061     \\
\textbf{Geography}                 &       0.6167  &        0.471     &     1.308  &         0.191        &       -0.307    &        1.541     \\
\textbf{History}                   &      -0.0161  &        1.134     &    -0.014  &         0.989        &       -2.239    &        2.207     \\
\textbf{Mathematics}               &      -0.2700  &        0.933     &    -0.289  &         0.772        &       -2.098    &        1.558     \\
\textbf{Medicine}                  &       1.8011  &        0.299     &     6.023  &         0.000        &        1.215    &        2.387     \\
\textbf{Political\_science}        &       0.3283  &        0.299     &     1.099  &         0.272        &       -0.257    &        0.914     \\
\textbf{Psychology}                &       0.4982  &        0.300     &     1.662  &         0.096        &       -0.089    &        1.085     \\
\textbf{Sociology}                 &       0.2015  &        0.345     &     0.584  &         0.559        &       -0.475    &        0.878     \\
\hline
\textbf{  No. Observations:  } &     559  & \textbf{  Log-Likelihood:    } &  -1453.4 & \textbf{  Dispersion:    } &  1 \\
\textbf{  Df Model:          } &      20     & \textbf{  Df Residuals:      } &     538   \\
 \textbf{  Pearson chi2:      } &   640.    & \textbf{  Deviance:          } &     596.38 \\

\end{tabular}
\caption{the Negative Binomial Regression Model Detail: the Effects of Atypicality of Data Combinations and Topic Atypicality on 3-Year Citations for paper published before 1990.  }
\end{center}
\end{table*}

\FloatBarrier
\textbf{-Alternative Impact Quantification: We examined the impact of using atypical topic and atypical combinations of datasets on the likelihood of becoming top 5\% hit papers – publications that received citations within the top 5\% in our dataset.(Table 30)}

\FloatBarrier

\begin{table*}[ht]
\begin{center}
  \renewcommand{\tablename}{Table S}
\tiny 

\begin{tabular}{lcccccc}
\tiny 
\textbf{Dep. Variable:}  top 5\% hit paper (binary)  & \textbf{coef} & \textbf{std err} & \textbf{z} & \textbf{P$> |$z$|$} & \textbf{[0.025} & \textbf{0.975]}  \\
\hline
\hline
\textbf{Year\_bin[(1930, 1974]]} &     -25.6837  &     3.03e+04     &    -0.001  &         0.999        &    -5.95e+04    &     5.94e+04     \\
\textbf{Year\_bin[(1974, 1979]]} &     -26.7401  &     1.63e+04     &    -0.002  &         0.999        &     -3.2e+04    &      3.2e+04     \\
\textbf{Year\_bin[(1979, 1984]]} &      -8.0424  &        1.114     &    -7.219  &         0.000        &      -10.226    &       -5.859     \\
\textbf{Year\_bin[(1984, 1989]]} &      -8.9517  &        1.135     &    -7.888  &         0.000        &      -11.176    &       -6.727     \\
\textbf{Year\_bin[(1989, 1994]]} &      -7.6733  &        0.636     &   -12.058  &         0.000        &       -8.921    &       -6.426     \\
\textbf{Year\_bin[(1994, 1999]]} &      -7.1462  &        0.519     &   -13.775  &         0.000        &       -8.163    &       -6.129     \\
\textbf{Year\_bin[(1999, 2004]]} &      -6.7747  &        0.509     &   -13.304  &         0.000        &       -7.773    &       -5.777     \\
\textbf{Year\_bin[(2004, 2009]]} &      -6.8619  &        0.516     &   -13.303  &         0.000        &       -7.873    &       -5.851     \\
\textbf{Year\_bin[(2009, 2014]]} &      -6.9771  &        0.509     &   -13.710  &         0.000        &       -7.974    &       -5.980     \\
\textbf{Year\_bin[(2014, 2019]]} &      -7.3252  &        0.514     &   -14.258  &         0.000        &       -8.332    &       -6.318     \\
\textbf{Atypicality\_of\_datasets}    &       0.4257  &        0.082     &     5.213  &         0.000        &        0.266    &        0.586     \\
\textbf{Topic\_atypicality}           &      -0.0251  &        0.067     &    -0.376  &         0.707        &       -0.156    &        0.106     \\
\textbf{NumDatasets\_log}             &      -0.2921  &        0.112     &    -2.610  &         0.009        &       -0.511    &       -0.073     \\
\textbf{Paper\_novelty}               &       0.2851  &        0.074     &     3.828  &         0.000        &        0.139    &        0.431     \\
\textbf{Data\_use\_frequency\_log}    &       0.1824  &        0.053     &     3.439  &         0.001        &        0.078    &        0.286     \\
\textbf{NumAuthor\_log}               &       0.8282  &        0.098     &     8.425  &         0.000        &        0.636    &        1.021     \\
\textbf{AuthorExprience\_log}         &      -0.0645  &        0.031     &    -2.076  &         0.038        &       -0.125    &       -0.004     \\
\textbf{ImpactFactor\_log}            &       1.1367  &        0.095     &    11.908  &         0.000        &        0.950    &        1.324     \\
\textbf{Biology}                      &       1.6279  &        1.461     &     1.114  &         0.265        &       -1.237    &        4.492     \\
\textbf{Business}                     &      -0.8907  &        1.140     &    -0.781  &         0.435        &       -3.125    &        1.344     \\
\textbf{Chemistry}                    &       0.5752  &        1.447     &     0.398  &         0.691        &       -2.261    &        3.411     \\
\textbf{Computer\_science}            &      -0.8210  &        1.739     &    -0.472  &         0.637        &       -4.229    &        2.587     \\
\textbf{Economics}                    &       1.3186  &        0.561     &     2.349  &         0.019        &        0.218    &        2.419     \\
\textbf{Engineering}                  &      -1.6164  &        3.005     &    -0.538  &         0.591        &       -7.507    &        4.274     \\
\textbf{Environmental\_science}       &       0.0320  &        2.463     &     0.013  &         0.990        &       -4.796    &        4.860     \\
\textbf{Geography}                    &      -0.5516  &        0.862     &    -0.640  &         0.522        &       -2.241    &        1.138     \\
\textbf{History}                      &       2.0969  &        2.666     &     0.787  &         0.432        &       -3.128    &        7.322     \\
\textbf{Materials\_science}           &     -60.2374  &     3.69e+05     &    -0.000  &         1.000        &    -7.23e+05    &     7.23e+05     \\
\textbf{Mathematics}                  &      -0.0273  &        1.924     &    -0.014  &         0.989        &       -3.799    &        3.744     \\
\textbf{Medicine}                     &       1.4709  &        0.345     &     4.268  &         0.000        &        0.795    &        2.146     \\
\textbf{Philosophy}                   &     -52.7229  &     1.79e+05     &    -0.000  &         1.000        &    -3.52e+05    &     3.51e+05     \\
\textbf{Political\_science}           &      -1.8463  &        0.773     &    -2.389  &         0.017        &       -3.361    &       -0.331     \\
\textbf{Psychology}                   &       0.1831  &        0.395     &     0.464  &         0.643        &       -0.590    &        0.956     \\
\textbf{Sociology}                    &       1.9954  &        0.686     &     2.910  &         0.004        &        0.651    &        3.339     \\
\hline

\textbf{  No. Observations:  } &     8812      & \textbf{  Log-Likelihood:    } &  -1453.2     \\
\textbf{  Df Model:          } &       33     & \textbf{  Df Residuals:      } &     8778    \\
 \textbf{  Pearson chi2:      } &   1.29e+04  & \textbf{  Deviance:          } &    2906.5   \\
\end{tabular}
\caption{Logistic regression model: investigating the impact of atypicality of data combinations and topic atypicality on achieving top 5 percent hit paper status. This study serves as a robustness test. The hit paper variable is binary, with 1 indicating that the publication received citations in the top 5 percent among all the papers in our dataset, and 0 otherwise. }
\end{center}
\end{table*}

\newpage

\FloatBarrier
\newpage
\textbf{(4) What type of research teams combine atypical datasets: Impact of Team Size and Team experience on likelihood of using Data Combination  (Table 31-32)} 

\FloatBarrier

\begin{table*}[ht]
\begin{center}
  \renewcommand{\tablename}{Table S}
\tiny 

\begin{tabular}{lcccccc}
  \textbf{Dep. Variable:}  Using Data Combination (using multiple dataset)                                    & \textbf{coef} & \textbf{std err} & \textbf{t} & \textbf{P$> |$t$|$} & \textbf{[0.025} & \textbf{0.975]}  \\
\hline
\hline
\textbf{Intercept}                 &      -1.0502  &        0.049     &   -21.373  &         0.000        &       -1.147    &       -0.954     \\
\textbf{NumAuthor\_log}            &       0.0950  &        0.020     &     4.815  &         0.000        &        0.056    &        0.134     \\
\textbf{Data\_use\_frequency\_log} &      -0.0490  &        0.007     &    -6.907  &         0.000        &       -0.063    &       -0.035     \\
\textbf{ImpactFactor\_log}         &       0.1356  &        0.020     &     6.847  &         0.000        &        0.097    &        0.174     \\
\hline

\textbf{Model:}                    & Logit & \textbf{   Pseudo R-squ.:         } &         0.003104   \\
 \textbf{  Log-Likelihood:    } &     -18293.   & \textbf{  Method:       } &       MLE    \\
\textbf{No. Observations:}                 &        30479      & \textbf{  Df Residuals:               } &   30475                 \\

\end{tabular}

\caption{Logistic regression results on the effect of team size (number of authors) of a publication on the using multiple dataset (data combination). An increase in the number of authors is associated with a higher probability of using multiple dataset (data combination).}
\end{center}
\end{table*}

\begin{table*}[ht]
\begin{center}
  \renewcommand{\tablename}{Table S}
\tiny 

\begin{tabular}{lcccccc}
\textbf{Dep. Variable:} Using Data Combination (using multiple dataset)  & \textbf{coef} & \textbf{std err} & \textbf{t} & \textbf{P$> |$t$|$} & \textbf{[0.025} & \textbf{0.975]}  \\
\hline
\hline
\textbf{Intercept}                 &      -1.0241  &        0.069     &   -14.858  &         0.000        &       -1.159    &       -0.889     \\
\textbf{AuthorExprience\_log}      &       0.0034  &        0.007     &     0.521  &         0.602        &       -0.009    &        0.016     \\
\textbf{Data\_use\_frequency\_log} &      -0.0448  &        0.007     &    -6.336  &         0.000        &       -0.059    &       -0.031     \\
\textbf{ImpactFactor\_log}         &       0.1478  &        0.020     &     7.511  &         0.000        &        0.109    &        0.186     \\
\hline

\textbf{Model:}                    & Logit & \textbf{   Pseudo R-squ.:         } &         0.002480    \\
 \textbf{  Log-Likelihood:    } &      -18305.    & \textbf{  Method:       } &       MLE    \\
\textbf{No. Observations:}                 &        30479      & \textbf{  Df Residuals:               } &   30475                 \\

\end{tabular}

\caption{Logistic regression results on the effect of team experience (average citation of authors) of a publication on using multiple dataset (data combination). An increase in the number of authors is associated with a higher probability of using multiple dataset (data combination).}
\end{center}
\end{table*}

\FloatBarrier
\newpage
\textbf{- Impact of Team Size and Team experience on Atypicality of dataset combination (Table 33-34). }

\FloatBarrier

\begin{table*}[ht]
\begin{center}
  \renewcommand{\tablename}{Table S}
\tiny 

\begin{tabular}{lcccccc}
\textbf{Dep. Variable:} Atypicality of dataset combination    & \textbf{coef} & \textbf{std err} & \textbf{t} & \textbf{P$> |$t$|$} & \textbf{[0.025} & \textbf{0.975]}  \\
\hline
\hline
\textbf{Intercept}                 &      -1.3018  &        0.031     &   -41.381  &         0.000        &       -1.363    &       -1.240     \\
\textbf{NumAuthor\_log}            &      -0.1284  &        0.011     &   -11.545  &         0.000        &       -0.150    &       -0.107     \\
\textbf{NumDatasets\_log}          &       1.0395  &        0.009     &   111.159  &         0.000        &        1.021    &        1.058     \\
\textbf{Data\_use\_frequency\_log} &       0.0041  &        0.005     &     0.784  &         0.433        &       -0.006    &        0.014     \\
\textbf{ImpactFactor\_log}         &       0.0151  &        0.010     &     1.446  &         0.148        &       -0.005    &        0.036     \\

\hline

\textbf{Model:}                            &        OLS       & \textbf{  R-squared:         } &        0.586    \\

\textbf{  Log-Likelihood:    } &    -8638.2  & \textbf{  F-statistic:       } &       3129.    \\

\textbf{No. Observations:}                 &       8836     & \textbf{  AIC:               } &    1.729e+04   \\
                \\

\end{tabular}

\caption{OLS regression results on the effect of team experience (average citation of authors) of a publication on Atypicality of dataset combination. An increase in the team experience is associated with a lower chance of using atypical dataset combination.}
\end{center}
\end{table*}

\begin{table*}[ht]
\begin{center}
  \renewcommand{\tablename}{Table S}
\tiny 
\begin{tabular}{lcccccc}
\textbf{Dep. Variable:} Atypicality of dataset combination  & \textbf{coef} & \textbf{std err} & \textbf{t} & \textbf{P$> |$t$|$} & \textbf{[0.025} & \textbf{0.975]}  \\
\hline
\hline
\textbf{Intercept}                 &      -1.1900  &        0.040     &   -29.416  &         0.000        &       -1.269    &       -1.111     \\
\textbf{AuthorExprience\_log}      &      -0.0194  &        0.004     &    -5.530  &         0.000        &       -0.026    &       -0.013     \\
\textbf{NumDatasets\_log}          &       1.0338  &        0.009     &   110.104  &         0.000        &        1.015    &        1.052     \\
\textbf{Data\_use\_frequency\_log} &      -0.0094  &        0.005     &    -1.855  &         0.064        &       -0.019    &        0.001     \\
\textbf{ImpactFactor\_log}         &       0.0071  &        0.010     &     0.679  &         0.497        &       -0.013    &        0.028     \\
\hline
\textbf{Model:}                            &                       OLS                       & \textbf{  R-squared:     } &      0.582  \\
\textbf{  Log-Likelihood:    } &   -8689.1              & \textbf{  F-statistic:} &3068.  \\

\textbf{No. Observations:}                 &                        8836                     & \textbf{  AIC:               } &      1.739e+04 \\

\end{tabular}
\caption{OLS regression results on the effect of team experience (average citation of authors) of a publication on atypicality of dataset combinations. An increase in the team experience is associated with a lower atypicality of dataset combinations.}
\end{center}
\end{table*}

\newpage
\subsection*{5. Regression Equations}

We employ fixed-effect Negative Binomial models to quantify the relationship between the atypicality of data usage and scientific impact. These models control for confounds such as publication year, dataset use frequency, number of authors, author experience (measured by average citation count of authors in the targeted publication), number of datasets, estimated impact factor, and disciplines. Alternative measurements, null models, and analyses with different dataset samples further support our results (the results from these regressions are shown in Section 4).

The initial analysis, conducted using Equation~\ref{eq:nbreg1}, investigates the relationship between the variables \emph{$\text{V}^{\text{DataComb}}_i$} (using data combination) and $\text{Impact}_i$  (citation impact) with control variables $X$ (defined above). The results are shown in Figure 1(A).

\begin{equation}
\text{Impact}_i \sim \text{NegativeBinomial}(\text{V}^{\text{DataComb}}_i+\sum_k X_{ik})
\label{eq:nbreg1}
\end{equation}

The next analysis, conducted using Equation~\ref{eq:nbreg2}, investigates the relationship between the variable \emph{$\text{A}^{\text{Data}}_i$} (atpicality of datasets combinations) and the citation impact The results are shown in Figure 2(A). 

\begin{equation}
\text{Impact}_i \sim \text{NegativeBinomial}(\text{A}^{\text{Data}}_i+\sum_k X_{ik})
\label{eq:nbreg2}
\end{equation}

We then examine different approaches for defining the atypicality of dataset combinations. Utilizing Equation~\ref{eq:nbreg3}, we analyze the relationship between the variables \emph{$\text{A}^{\text{TopicA}}_i$} (topic atypicality) and citation impact. The results are shown in Figure 3(B).

\begin{equation}
\text{Impact}_i \sim \text{NegativeBinomial}(\text{A}^{\text{Data}}_i+\text{A}^{\text{TopicA}}_i+\sum_k X_{ik})
\label{eq:nbreg3}
\end{equation}



Finally, the relationship between $\text{V}^{\text{Teamsize}}$ (team size), $\text{V}^{\text{TeamExperience}}$ (team experience), \emph{$\text{V}^{\text{DataComb}}_i$} (data combination), $\text{A}^{\text{Data}}_i$ is examined using Equation~\ref{eq:nbreg5}, \ref{eq:nbreg6}, \ref{eq:nbreg7}, and \ref{eq:nbreg8}. The results are shown in Figures 4(A) and 4(B).

\begin{equation}
\emph{$\text{V}^{\text{DataComb}}_i$} \sim \text{Logistic}(\text{V}^{\text{Teamsize}}_i+\sum_k X_{ik})
\label{eq:nbreg5}
\end{equation}

\begin{equation}
\emph{$\text{V}^{\text{DataComb}}_i$} \sim \text{Logistic}(\text{V}^{\text{TeamExperience}}_i+\sum_k X_{ik})
\label{eq:nbreg6}
\end{equation}

\begin{equation}
\text{A}^{\text{Data}}_i \sim \text{OLS}\text{(V}^{\text{Teamsize}}_i+\sum_k X_{ik})
\label{eq:nbreg7}
\end{equation}

\begin{equation}
\text{A}^{\text{Data}}_i \sim \text{OLS}\text{(V}^{\text{TeamExperience}}_i+\sum_k X_{ik})
\label{eq:nbreg8}
\end{equation}
 

\subsection*{6. Robustness Check --- Matching}

(1) {\bf The effect of dataset combinations on scientific impact}  We conducted additional and more intuitive robustness tests using a matching method. For each paper that uses exactly two datasets, we identified a paper published in the nearest year that uses only one of the two datasets (randomly selecting the matched paper if there were multiple possible matches). We first conducted a paired t-test to compare the distribution of 3, 5, and 10-year citations for papers that use exactly two datasets and that of their matched set. The mean number of citations of papers that use two datasets is significantly larger (P $<$ 0.05). Second, we find the ratios of citations between the papers using two datasets and papers using one dataset have a median of 1.11,1.13,1.14, and a mean of 2.75, 3.07, and 3.5 for 3, 5, and 10-year citations, respectively \footnote{Since we cannot calculate the mean value of the ratio when the year citation count of paper using one dataset is zero due to the resulting infinite value, we exclude these cases when computing the mean and distribution of the ratio. We do not exclude these observations when performing other analyses, such as computing the median or performing the paired t-tests}. Additionally, we used a one-sample t-test to determine if the mean is significantly different from 1 and a sign test to assess if the median is significantly different from 1, both of which yielded significant results (P $<$ 0.05). All results indicate that papers using two datasets, on average, receive significantly more citations than those using only one dataset.
 Our results remain significant and consistent with the regression results and we have attached the distribution of the ratio in Figure S4.

\begin{figure}[htbp]
   \renewcommand{\figurename}{Figure S}
  \centering
  \includegraphics[width=0.8\textwidth]{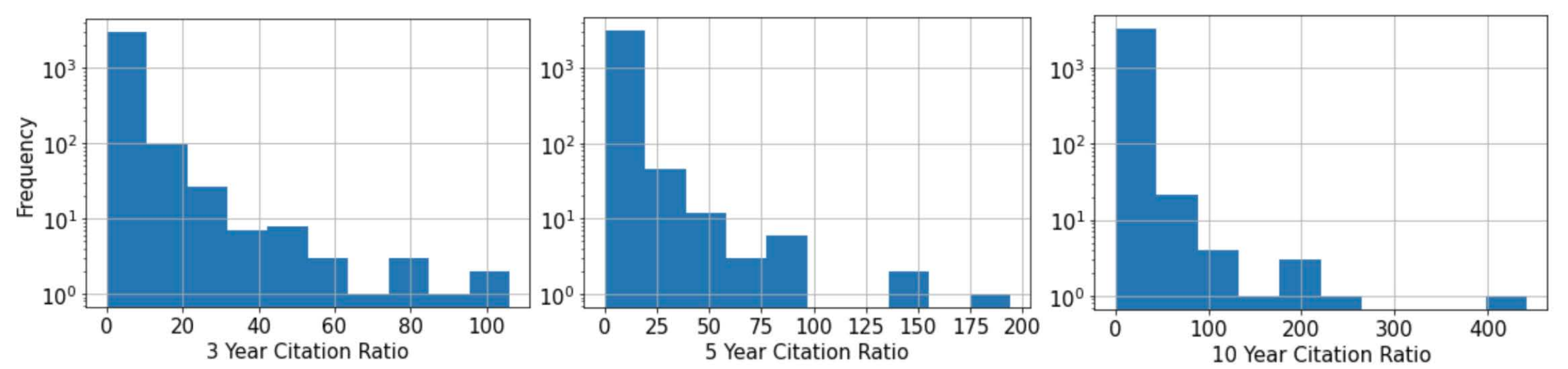}
  \caption{Distribution of citation ratio for number of data matching test}
  \label{fig:numdatamatch}
\end{figure}

(2) {\bf Atypical combinations of datasets associate with high impact}
Similar to the prior analysis, we also conducted an additional and more intuitive robustness test using a matching method. We first selected papers that use exactly two datasets and identified those with a top 10\% atypicality score in that group (Mean atypicality = 0.49). We then matched each selected focal paper with another paper that uses one of the two datasets used by the focal paper but has the lowest atypicality score among such papers, filtering out pairs where the matched paper has an equal or higher atypicality score than the focal paper (Mean atypicality = 0.41). We first conducted a paired t-test to compare the distribution of 3, 5, and 10-year citations for papers with high and low atypicality and found the mean number of citations of papers with high atypicality is significantly larger (P $<$ 0.05). Then, we find the citation ratios of papers with high atypicality scores compared to their matched counterparts with low atypicality scores have median values of 1.25, 1.26, and 1.42 for the 3, 5, and 10-year citations, respectively. Correspondingly, the mean values are 3.61, 4.36, and 6.03 \footnote{Since we cannot calculate the mean value of the ratio when the year citation count of the low atypicality paper is zero due to the resulting infinite value, we exclude these cases to compute the mean and distribution of the ratios. We do not exclude these observations when performing other analyses, such as computing the median or performing the paired t-tests.}. Additionally, we used a one-sample t-test and a sign test to determine if the mean and median were significantly different from 1, both of which yielded significant results (P $<$ 0.05). All results indicate that papers using two datasets with a higher atypicality score, on average, receive significantly more citations than those using only one dataset. Our results are consistent with the regression results and we have attached the distribution of the ratio in Figure S5.


\begin{figure}[htbp]
   \renewcommand{\figurename}{Figure S}
  \centering
  \includegraphics[width=0.8\textwidth]{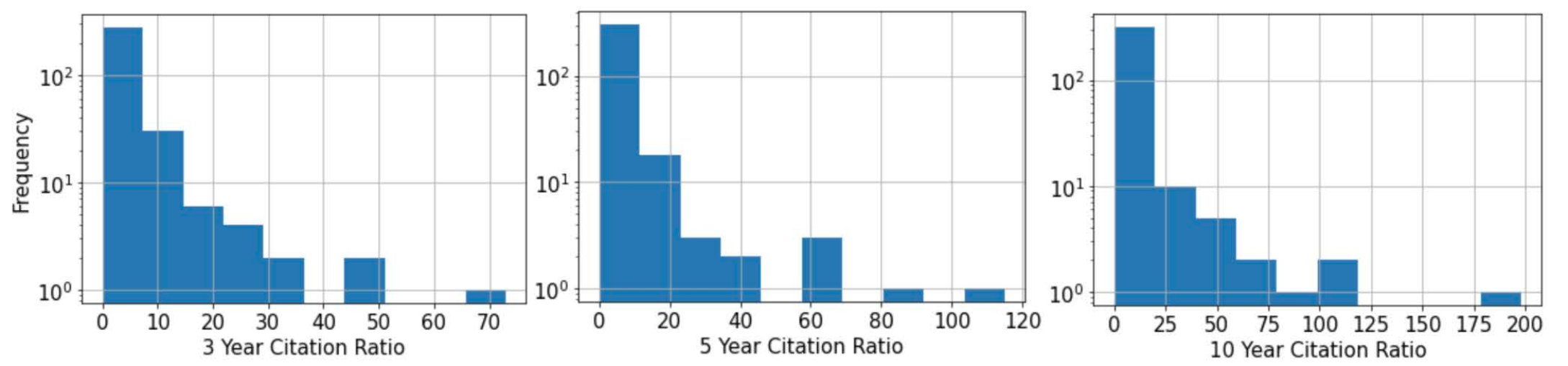}
  \caption{Distribution of citation ratio for atypicality matching test}
  \label{fig:numdatamatch}
\end{figure}

\bibliographystyle{unsrt}
\end{document}